\documentclass[twocolumn, amssymb,nobibnotes,aps,prb,nopacs]{revtex4}

\usepackage{amsmath}
\usepackage{graphicx}
\usepackage{subfig}
\usepackage[justification=RaggedRight,font=footnotesize]{caption}

\usepackage{multirow}

\begin{document}

\title{Simulation study of overtaking of ion-acoustic solitons 
\\ in the fully kinetic regime}

\author{S. M. Hosseini Jenab\footnote{Email: Mehdi.Jenab@nwu.ac.za}}
\author{F. Spanier \footnote{Email: Felix@fspanier.de}}
\affiliation{Centre for Space Research, North-West University, Potchefstroom Campus,
Private Bag X6001, Potchefstroom 2520, South Africa}

\date{\today}

\begin{abstract}
The overtaking collisions of ion-acoustic solitons (IASs)
in presence of trapping effects of electrons are studied 
based on a fully kinetic simulation approach. 
The method is able to provide all the kinetic details of the 
process alongside the fluid-level quantities self consistently. 
Solitons are produced naturally by utilizing the chain formation phenomenon,
then are arranged in a new simulation box to test different scenarios of overtaking collisions.
Three achievements are reported here.
Firstly, simulations prove the long-time life span of the ion-acoustic solitons 
in the presence of trapping effect of electrons (kinetic effects), 
which serves as the benchmark of the simulation code.
Secondly, their stability against overtaking mutual collisions is established 
by creating collisions between solitons with different number and shapes of trapped electrons,
i.e. different trapping parameter.
Finally, details of solitons during collisions for both ions and electrons are provided
on both fluid and kinetic levels.
These results show that on the kinetic level,
trapped electron population accompanying each of the solitons are exchanged between the solitons
during the collision. 
Furthermore, the behavior of electron holes accompanying solitons contradicts 
the theory about the electron holes interaction developed based on kinetic theory. 
They also show behaviors much different from other electron holes witnessed in 
processes such as nonlinear Landau damping
(Bernstein-Greene-Kruskal -BGK- modes) or beam-plasma interaction (like two-beam instability).
\end{abstract}
\maketitle

\section{Introduction} \label{Sec_Intro}
Solitons are defined as nonlinear localized solutions which are stable against mutual collisions,
namely head-on and overtaking collisions\cite{Wadati2001841}. 
These structures emerge after the collisions without any changes in their physical features
(such as velocity, height, width and shape in both velocity and spatial directions). 
Stability is defined as their physical features remaining the same before and after the collisions
(and not during collisions). 
As far as physical features are considered, solitons act as pseudo-particles, 
hence the suffix ``ton'' in the word ``soliton''\cite{Zabusky1965}.  
During collisions, however, features of two solitons participating in collision merge and overlap. 
Hence, a collision between two solitons is interpreted as their overlapping in spatial direction.
Solitons can be regarded as a subclass of solitary waves, nonlinear localized structures 
propagating steadily in a dissipative medium. 
``Stability against mutual collisions'' distinguishes the two concepts apart. 
Although in the context of plasma physics, especially for ion-acoustic solitons (IASs),
these two terms have been used interchangeably.

Historically, solitons were discovered in context of plasma fluid simulation\cite{Zabusky1965}, 
and since then 
theoretical and simulation approaches based on ``fluid framework''
have played a major role in studying different forms of solitons
in plasma physics\cite{kuznetsov1986soliton,shafranov2012reviews}.
In the fluid framework, densities of plasma species are considered as the their physical features. 
Temporal evolutions of these densities, coupled by Poisson's equation, produce the whole physical picture
of solitons by providing the electric potential and field for any given time. 
However, densities are reduced forms of distribution functions and hence kinetic effects, 
related to distribution functions, 
stays beyond the scope of fluid framework. 
In the experimental studies, the kinetic details are either not reported or overlooked
\cite{ikezi1970formation,lonngren1983soliton,tran1979ion}.  
Theoretical approaches such as the Sagdeev pseudo-potential\cite{Sagdeev} 
and the BGK method\cite{bernstein1957exact} supply a chance 
to incorporate kinetic effects into solitons studies. 
However, this comes at the price of losing the temporal evolution, 
which consequently means that stability against mutual collisions can not be studied. 
These methods are able to discover the nonlinear solutions, but can't prove/disprove them as solitons.

To address these limitations in the studies of solitons, we have employed a fully kinetic simulation approach, 
which can encompass the kinetic effects and supply the temporal evolution of the physical features. 
In this simulation method the dynamics of the plasma species, e.g. electrons and ions, 
are followed based on solving the Vlasov equations. 
Therefore, the dynamics of distribution functions, kinetic details, are provided 
and the fluid-level quantities,
i.e. densities of each species, are achieved self-consistently.
This theoretical framework, removes the limitations faced in the previous studies 
such as
small-amplitude limitation of reductive perturbation method i.e. KdV model,
the absence of temporal evolution in the Sagdeev's approach or the BGK method
and the lack of kinetic effects in fluid model.

One of the major kinetic effects in the context of solitons in plasmas is called \textit{trapping effect}. 
Particles in a certain range of energy resonate with the potential well imposed by a soliton, 
and oscillate inside the well, hence are trapped by the soliton. 
Schamel has utilized a self-developed version of the BGK method\cite{schamel_1} to integrate the trapping effect
into the study of ion-acoustic solitons (IASs). 
Trapping effect (controlled by the trapping parameter $\beta$) introduces its own extra nonlinearity to the KdV equation. 
Comparing this new nonlinearity with usual nonlinearity of KdV equation results in three regimes\cite{schamel_3}, 
each having their own 
fluid dynamical equations, namely KdV, Schamel-KdV and Schamel equations.
However it remains inconclusive, 
if the solutions should be regarded as solitons, 
i.e. if they can survive mutual collisions, 
due to the absence of the temporal evolution in the BGK method. 
Different fluid-based simulation methods have been employed to respond this concern\cite{Kakad2013,Sharma2015}. 
However, their results are limited since trapping effect can't be comprehensively 
considered by fluid models specially in case of large-amplitude solitons\cite{Kakad20145589}. 
Moreover, Particle-in-cell (PIC) simulation methods suffer from their inherit noise level
and can't provide a clear view of the kinetic-level interactions\cite{Kakad20145589,Qi20153815}.

Here, our main focus is on the overtaking collisions of ion-acoustic solitons 
in presence of electron trapping in two-species plasmas.
However, we need to show that simulation method is adjusted and fine-tunned 
for long-time nonlinear simulations.
Therefore, firstly the results of IASs propagation for long-time runs are presented in Sec.\ref{SubSec_propagation}. 
This section stands as the benchmarking of the simulation code which can assure 
the reliability of the simulation results in the long-time nonlinear stage. 
Afterwards, in Sec.\ref{SubSec_Stability} the question 
about the stability of IASs against collisions (here overtaking collisions)
is addressed.
Collisions between solitons with different sizes and trapping parameter
are presented covering different regimes proposed by Schamel. 
Note that interaction time of head-on collisions are shorter than of overtaking ones. 
Hence stability against overtaking collisions should be regarded as the ultimate test for the stability.

Furthermore, the details of a mutual overtaking collision between two IASs is presented in the phase space by 
showing the temporal evolution of distribution function of both species (Sec.\ref{SubSec_During}). 
In case of electron distribution function, kinetic details of two electron holes accompanying 
IASs during overtaking collisions are studied alongside other types of shapes.
The interaction between the holes reveals two contradictions with the existing theories.
Firstly, the results unveil the complexity of 
interaction of solitons on kinetic level, i.e. particle exchanging,
in contrast to the simple comprehension provided by the fluid framework. 
Secondly, it shows that the electron holes don't merge while exchanging trapped particles. 
Electron holes (trapped electrons population) has shown a tendency of merging in context of other 
phenomena in plasmas such as BGK modes, two stream instability. 
Our results show a very different tendency among the electron holes 
when they are accompanying IASs. 

\section{Basic Equations and Numerical Scheme} \label{Sec_Model}
Equations and quantities are normalized based on table \ref{table_normalization}. 
Hence the normalized Vlasov-Poisson set of equations reads as follow:
\begin{multline}
\frac{\partial f_s(x,v,t)}{\partial t} 
+ v \frac{\partial f_s(x,v,t)}{\partial x} 
\\ +  \frac{q_s}{m_s} E(x,t) \frac{\partial f_s(x,v,t)}{\partial v} 
= 0, \ \ \  s = i,e
\label{Vlasov}
\end{multline}
\begin{equation}
\frac{\partial^2 \phi(x,t)}{\partial x^2}  = n_e(x,t) - n_i(x,t)
\label{Poisson}
\end{equation}
where $s = i,e$ represents the corresponding species.
They are coupled by density integrations for each species to form a closed set of equations:
\begin{align}
n_s(x,t) &= n_{0s} N_s(x,t) \\
N_s(x,t) &= \int f_s(x,v,t) dv
\label{density}
\end{align}
in which $N$ stands for the number density.
Note that by this normalization,
ion sound velocity and electron plasma frequency are 
$v_C = 8.06$ and $\omega_{pe} = 10.0$, respectively.

\begin{table}
\small
\caption{Normalization of quantities.}
\begin{ruledtabular}
\begin{tabular}{cccc}
   \multirow{2}{*}{Name}&   \multirow{2}{*}{Symbol}& 
   \multicolumn{2}{c}{Normalized by}  \\
  {}&  {}&  Name& formula \\
 \hline
  Time     & $\tau$  	&ion plasma frequency 		&$\omega_{pi}  = {\big(\frac{n_{i0} e^2}{m_i \epsilon_0}\big)^{\frac{1}{2}} }$   \\
  Length   & $L$  	&ion Debye length		&$\lambda_{Di} = \sqrt{ \frac{\epsilon_0 K_B T_i}{n_{i0} e^2}   }$   \\
  Velocity & $v$  	&ion thermal velocity		&$v_{th_i} = \sqrt{\frac{K_B T_i}{m_i}}$   \\
  Energy   & $E$  	&{-------}			&$K_B T_i$   \\
  Potential   & $\phi$  	&{-------}		&$\frac{K_B T_i}{e}$   \\
  Charge 		&$q$		&elementary charge &$e$ \\
  Mass 			&$m$		&ion mass		&$m_i$
\end{tabular}
\end{ruledtabular}
\label{table_normalization}
\end{table} 

The Schamel distribution function \cite{schamel_1} has been utilized
as the initial distribution function to 
invoke a self-consistent hole in phase space and a localized compressional density profile in density. 
The normalized version of it reads as follow:
\begingroup\makeatletter\def\f@size{8.3}\check@mathfonts
\def\maketag@@@#1{\hbox{\m@th\large\normalfont#1}}%
\begin{equation*}
f_{s}(v) =  
  \left\{\begin{array}{lr}
     A \ exp \Big[- \big(\sqrt{\frac{\xi_s}{2}} v_0 + \sqrt{E(v)} \big)^2 \Big]   &\textrm{if}
      \left\{\begin{array}{lr}
      v<v_0 - \sqrt{\frac{2E_{\phi}}{m_s}}\\
      v>v_0+\sqrt{\frac{2E_{\phi}}{m_s}} 
      \end{array}\right. \\
     A \ exp \Big[- \big(\frac{\xi_s}{2} v_0^2 + \beta_s E(v) \big) \Big] &\textrm{if}  
     \left\{\begin{array}{lr}
      v>v_0-\sqrt{\frac{2E_{\phi}}{m_s}} \\
      v<v_0 + \sqrt{\frac{2E_{\phi}}{m_s}} 
      \end{array}\right.
\end{array}\right.
\label{Schamel_Dif}
\end{equation*}\endgroup
in which $A = \sqrt{ \frac{\xi_s}{2 \pi}} n_{0s}$,
and $\xi_s = \frac{m_s}{T_s}$ are the amplitude and the normalization factor respectively.
$E(v) = \frac{\xi_s}{2}(v-v_0)^2 + \phi\frac{1}{T_s q_s}$ 
represents the (normalized) energy of particles.
$v_0 = 0$ stands for the velocity of the initial density perturbation (IDP).
\begin{figure}
  \subfloat{\includegraphics[width=0.5\textwidth]{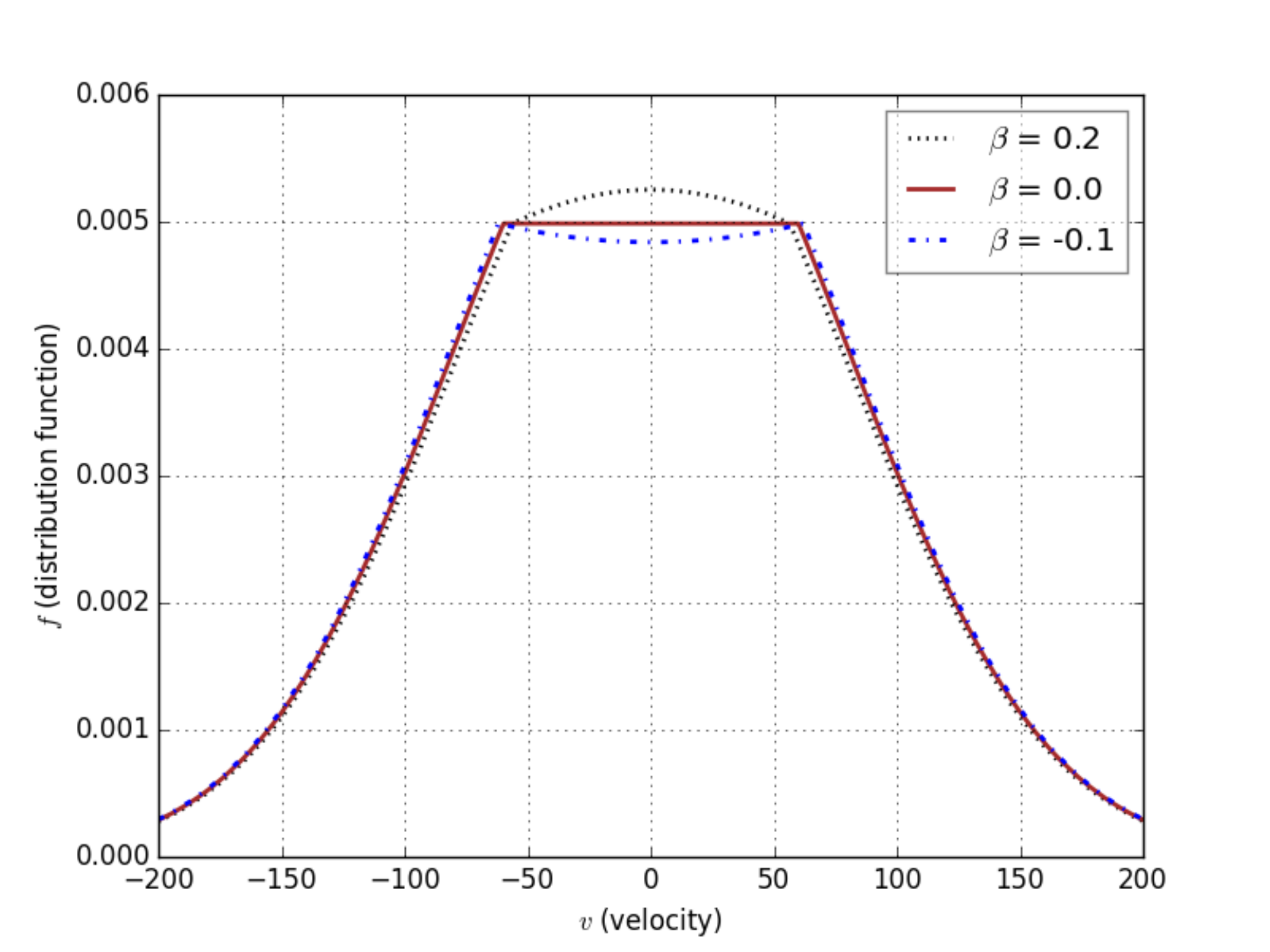}}
  \caption{Trapped electrons distribution function around $v_0=0$ 
  appears as a hole ($\beta<0$), a plateau ($\beta=0$) and a hump($\beta>0$)
  in the velocity direction.}
  \label{DF_Beta}
\end{figure}
$\beta$ is the so called \textit{trapping parameter} 
which describes the distribution function of trapped particles
around $v_0$. 
Based on $\beta$, Fig. \ref{DF_Beta} shows that the distribution function of trapped particles 
can take three different types of shapes, 
namely \textit{hole} ($\beta<0$), \textit{plateau} ($\beta = 0$) and \textit{hump} ($\beta>0$).

The hole in phase space produces a compressional pulse in the number density 
which is in turn introduces a localized structure in the 
electric potential.
Early in temporal evolution, the initial density perturbation (IDP) breaks 
into two oppositely moving density perturbations (MDP) due to the symmetry in the 
velocity direction of the distribution function \cite{jenab2016IASWs}. 
Then these MDPs split into number of IASs through the chain formation process. 
Note that the resulted IASs are not mathematical structures 
imposed to the system and are produced self-consistently. 
Then, these solitons are isolated and inserted 
into new simulation boxes to create different scenarios of overtaking collisions. 
In other words, the distribution function of these self-consistent solitons 
are inserted into certain places in the spatial direction, 
while the rest of the new simulation box is filled by 
the unperturbed Maxwellian distribution function:
$ f_m = \sqrt{ \frac{1}{2 \pi} \exp \big( - \frac{v^2}{2} \big)}.$

The constant parameters which remain fixed through all
of our simulations include: 
mass ratio $\frac{m_i}{m_e} = 100$,
temperature ratio $\frac{T_e}{T_i} = 64$
and $L = 1024$, 
where L is the length of the simulation box.
The periodic boundary condition is adopted on the spatial direction. 

The kinetic simulation approach utilized here 
is based on the Vlasov-Hybrid Simulation (VHS) method in which 
a distribution function is modeled by phase points\cite{nunn1993novel,kazeminezhad2003vlasov,jenab2011preventing}. 
The arrangement of phase points in the phase space 
at each time step provides the distribution function,
and hence all the kinetic momentums e.g. density, entropy and etc. 
The initial value of distribution function associated to each of the phase points stay intact
during simulation which guarantees the positiveness of distribution function
under any circumstances. 
Deviation of the conservation laws, e.g. conservation of entropy and energy, 
are closely monitored to stay below one percent. 

Each time step routine of the simulation consists of three steps which are summarized below. 
First step includes integrating distribution functions to achieve number densities. 
Then, by plugging these densities into Poisson's equation, electric potential are obtained. 
Poisson's equation is solved here based on a parallelized multi-grid method. 
On the third step, the Vlasov equation for each species is solved based on 
characteristics method utilizing leap-frog scheme to find the 
new arrangement of the phase points in the phase space for the next step. 

\section{Results and Discussion} \label{Sec_Results}
\subsection{Long time propagation of IA solitons} \label{SubSec_propagation}
\begin{figure}
  \subfloat{\includegraphics[width=0.5\textwidth]{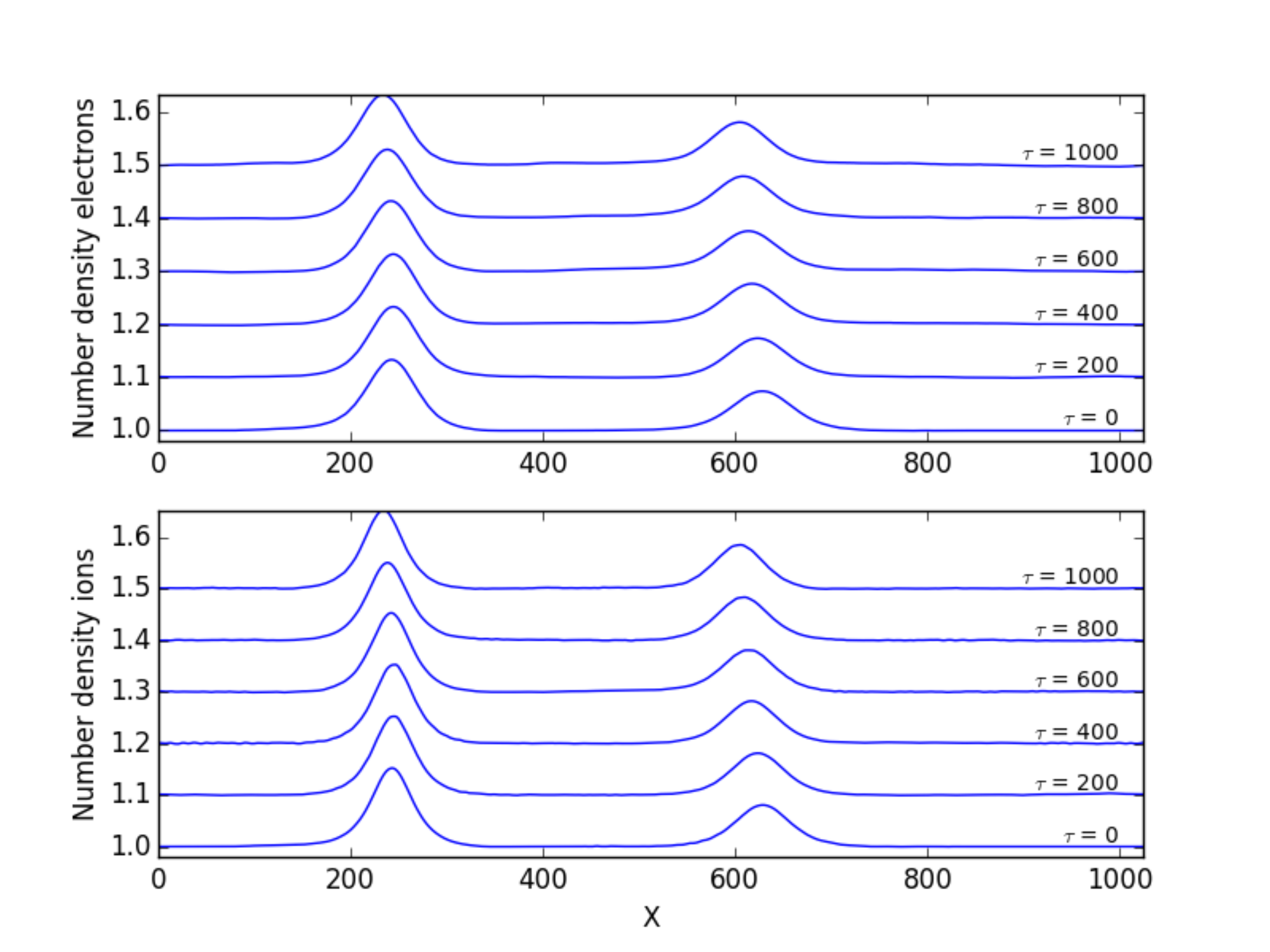}}
  \caption{Temporal evolution of electrons/ions number density is shown from $\tau = 0$ until $\tau = 1000$
  for two IASs with different trapping parameters, namely $\beta = 0.2$ and $\beta = 0$ 
  which are placed on left and right side of plot respectively.
  These two solitons have nearly the same velocity \cite{jenab2016IASWs} and 
  hence don't collide. Note that solitons are shown 
  in the frame which moves with the their average velocity ($v=9.1$). 
  The long-time propagation of IASs proves the reliability of the simulation code and serve as a benchmark.}
  \label{Fig_PlR_ZsR_num}
\end{figure}

\begin{figure}
 \begin{tabular}{c c}
  \ \ \subfloat{\includegraphics[width=0.26\textwidth]{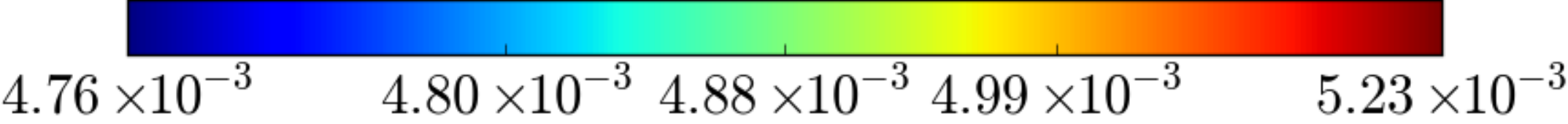}} &
  \ \ \ \subfloat{\includegraphics[width=0.26\textwidth]{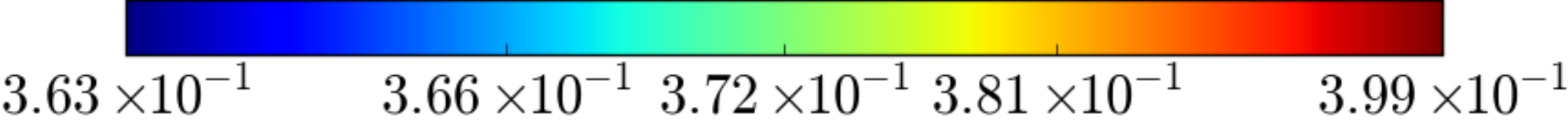}}\\
    \subfloat{\includegraphics[width=0.25\textwidth]{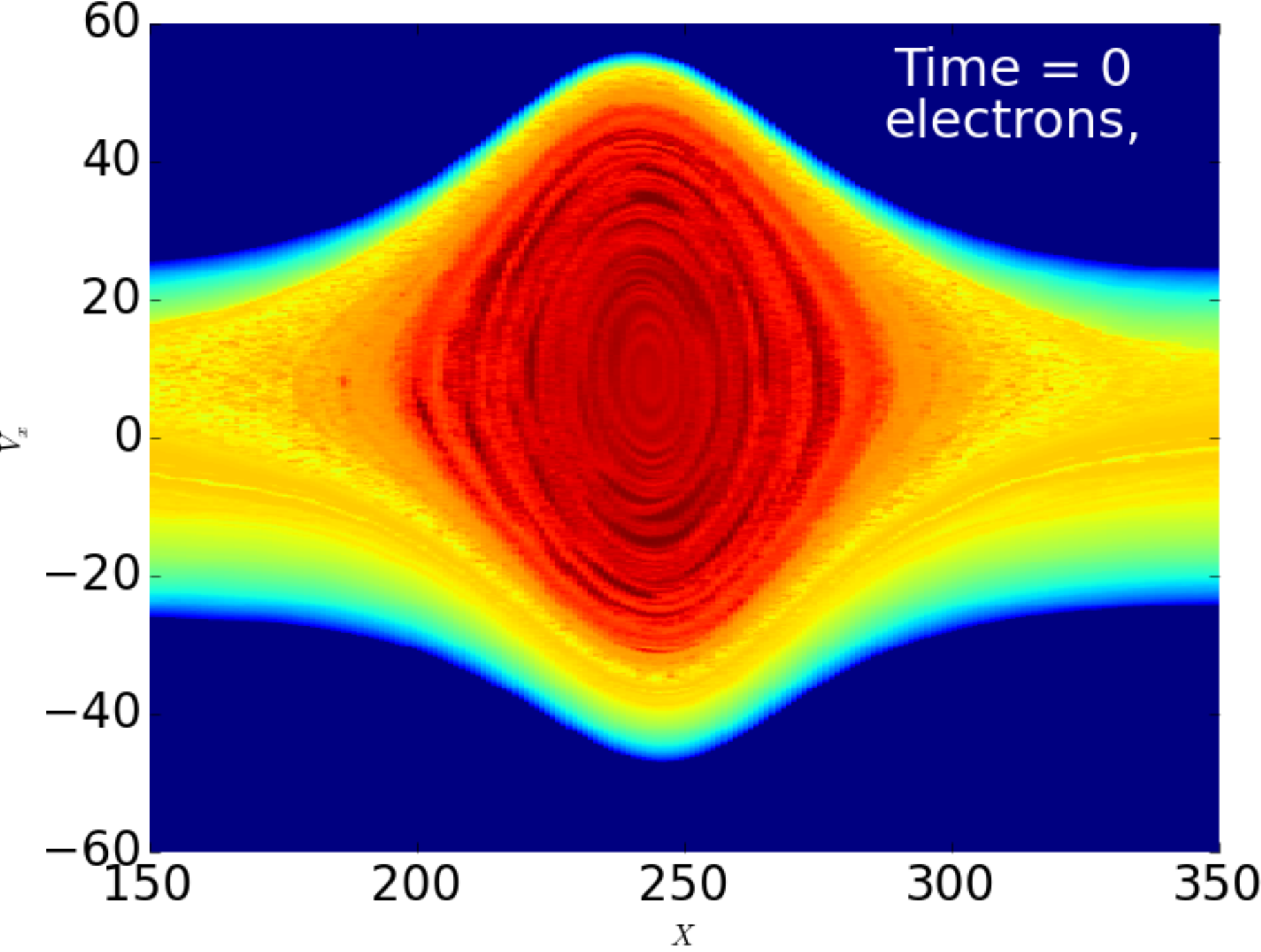}}&
  \subfloat{\includegraphics[width=0.25\textwidth]{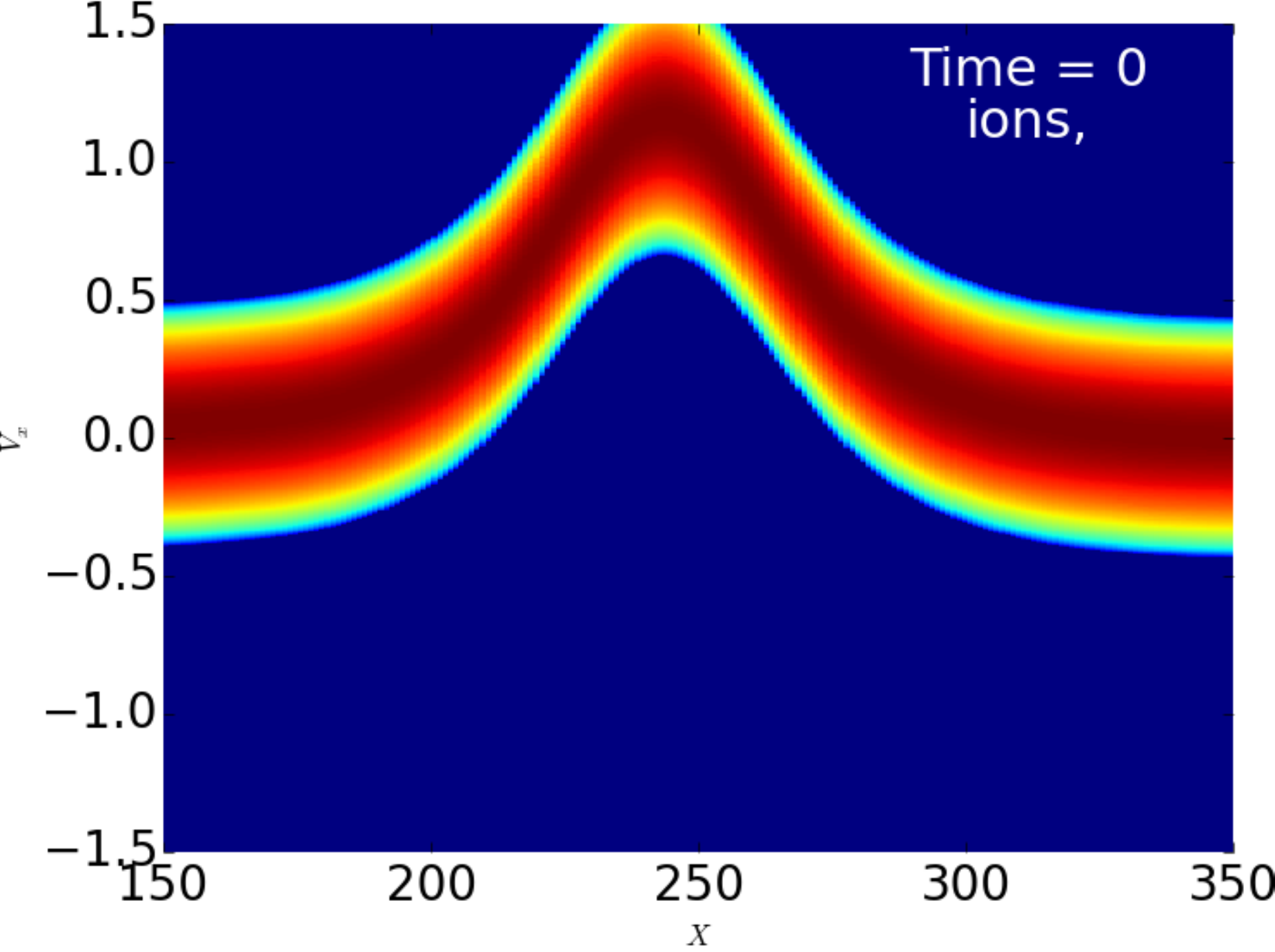}}\\
  \subfloat{\includegraphics[width=0.25\textwidth]{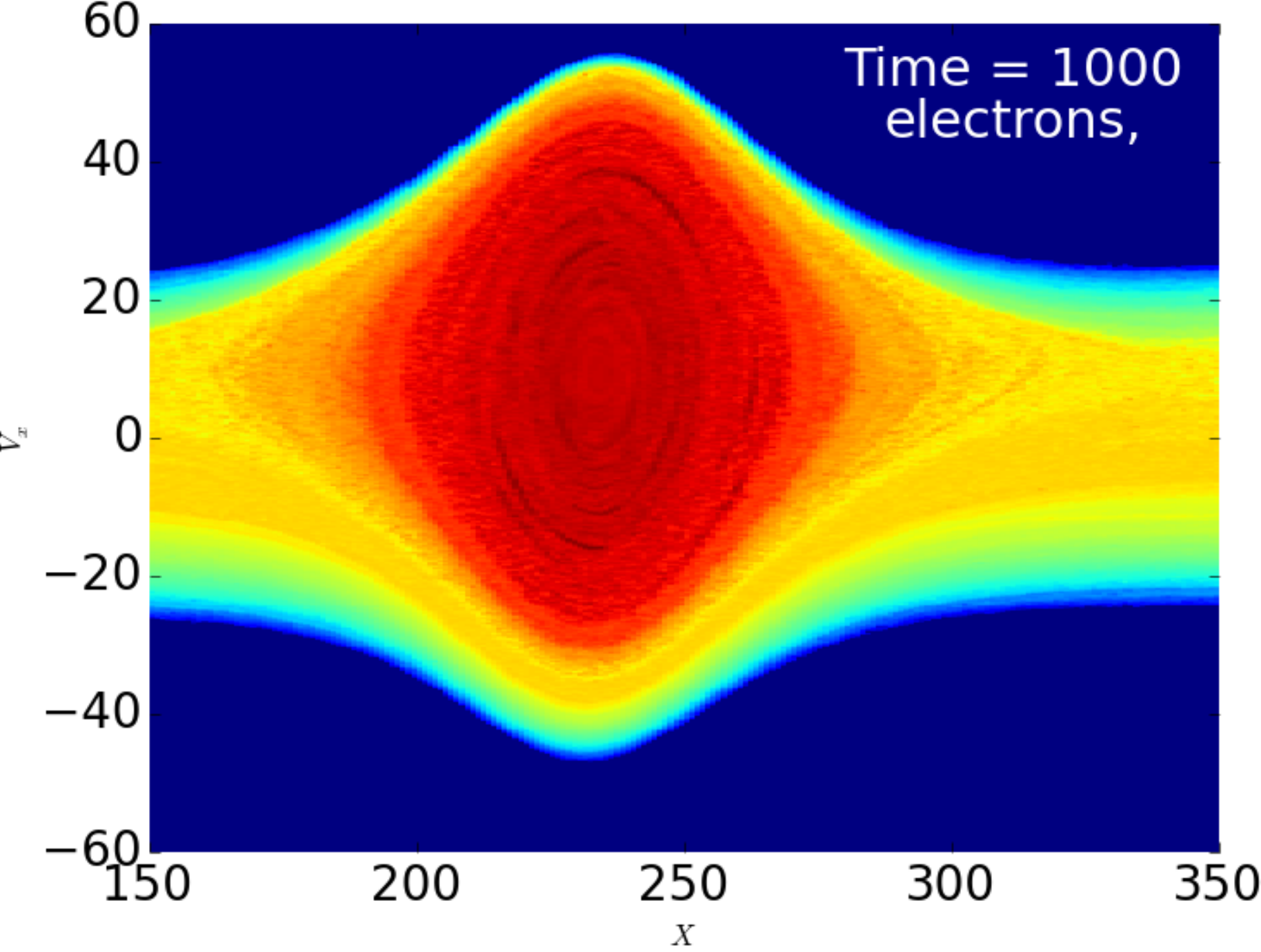}} &
  \subfloat{\includegraphics[width=0.25\textwidth]{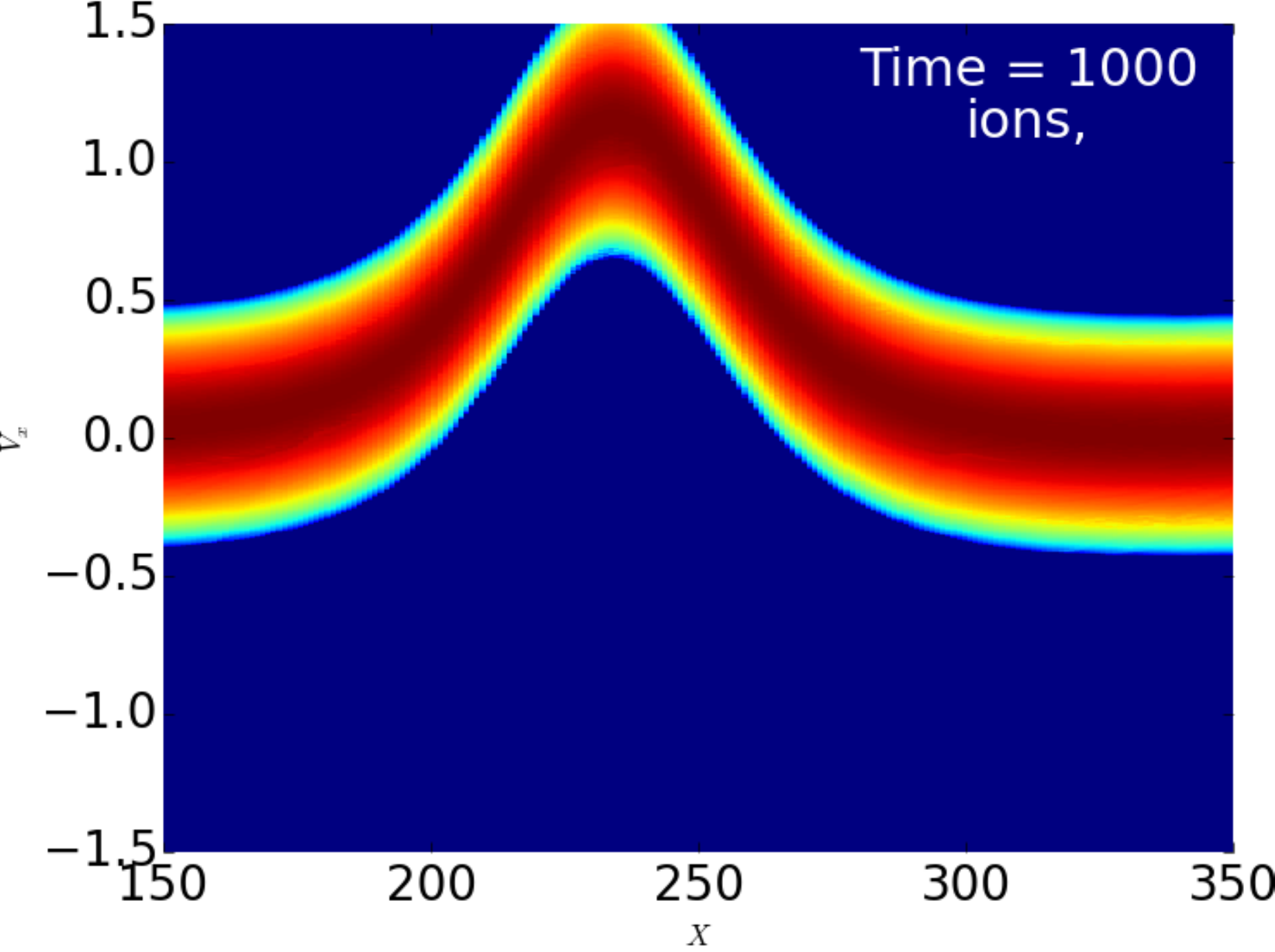}}\\
  \multicolumn{2}{c}{(a) the soliton with $\beta = 0.2$}\\
  \multicolumn{2}{c}{}\\ \multicolumn{2}{c}{}\\
 \end{tabular}
 \begin{tabular}{cc}
  \ \ \subfloat{\includegraphics[width=0.26\textwidth]{3_c_1_PlR_ZsR_colorbar_electrons.pdf}}&
  \ \ \ \subfloat{\includegraphics[width=0.26\textwidth]{3_c_2_PlR_ZsR_colorbar_ions.pdf}}\\
  
  \subfloat{\includegraphics[width=0.25\textwidth]{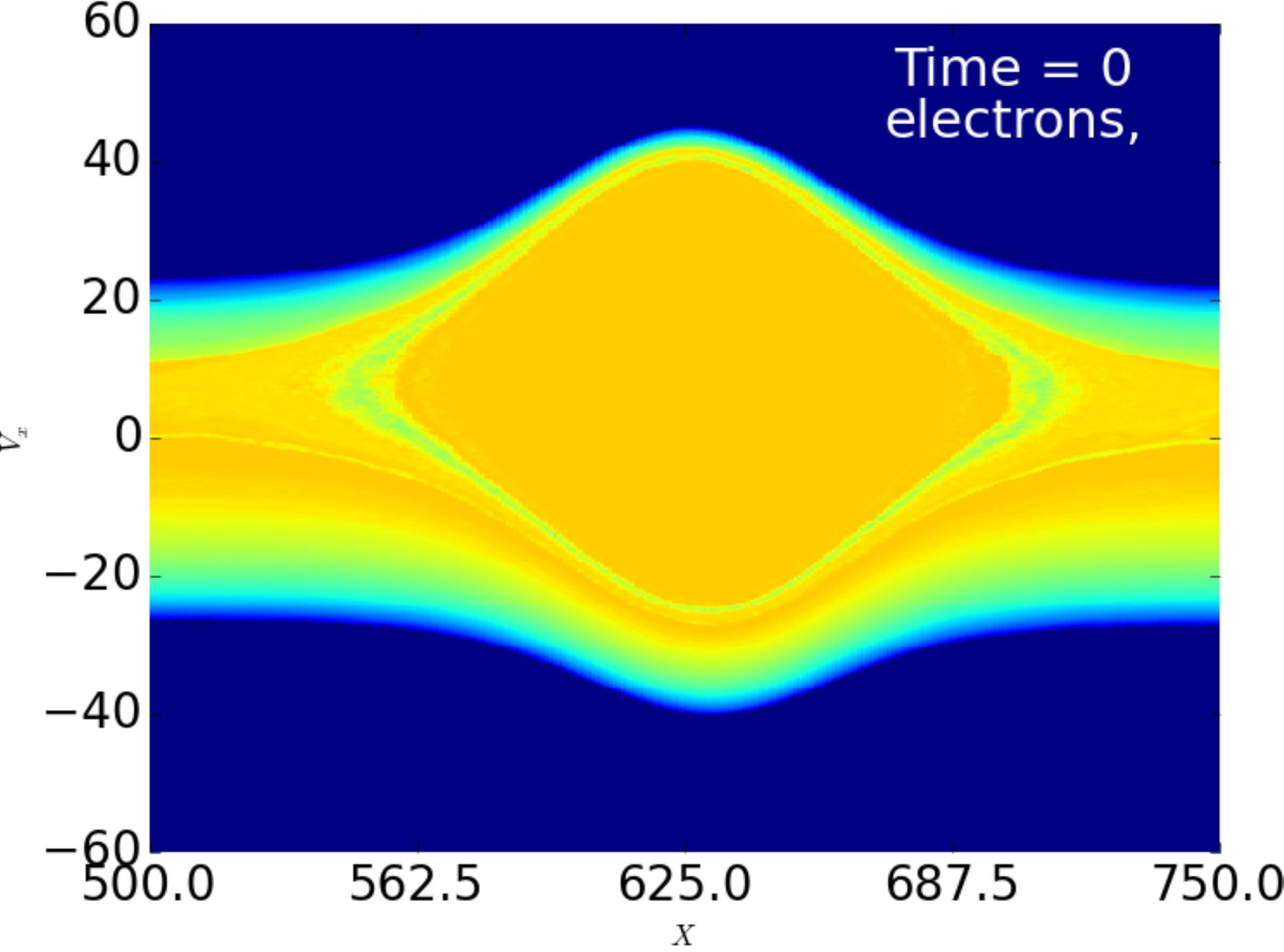}}&
  \subfloat{\includegraphics[width=0.25\textwidth]{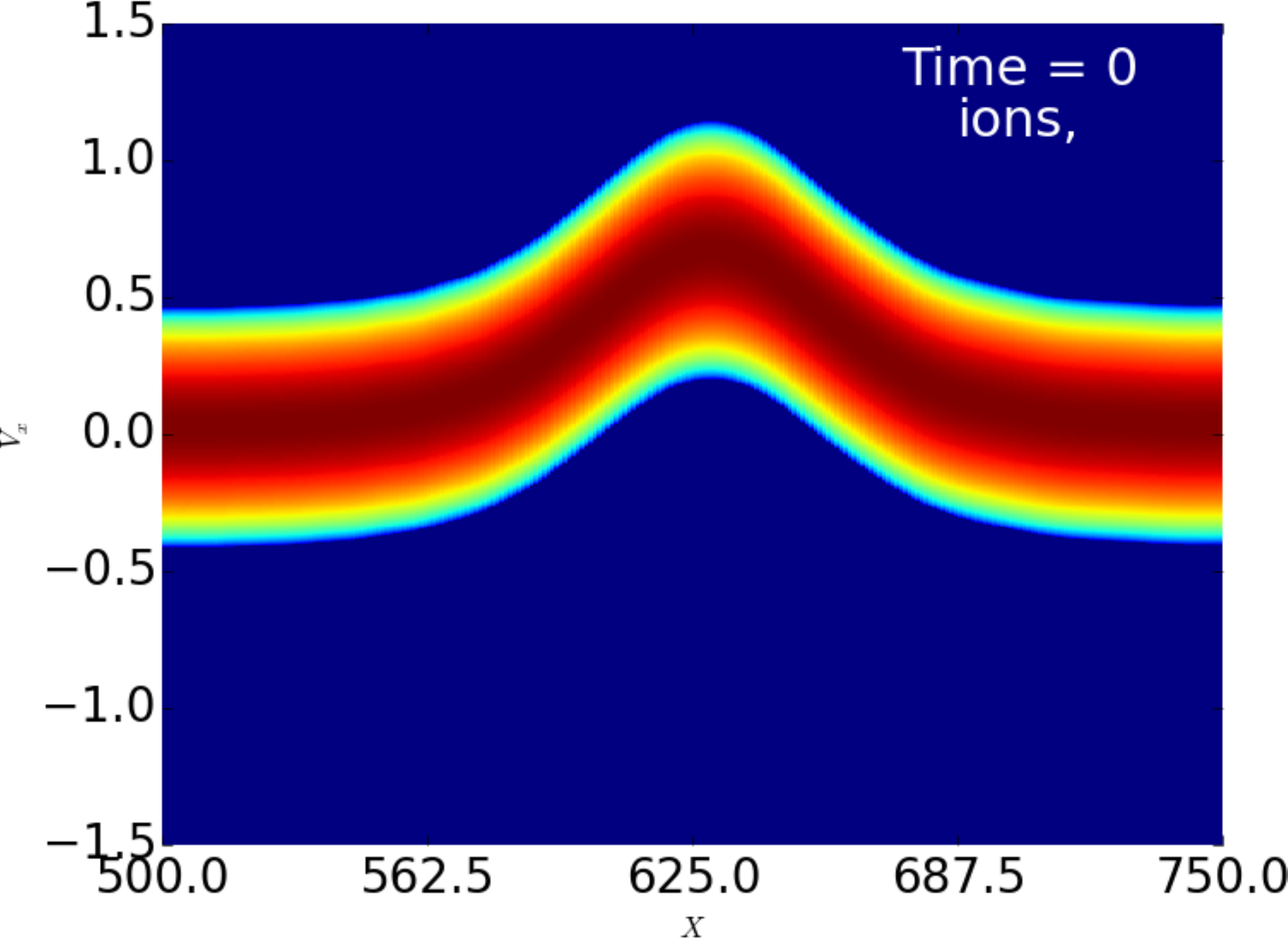}}\\
  \subfloat{\includegraphics[width=0.25\textwidth]{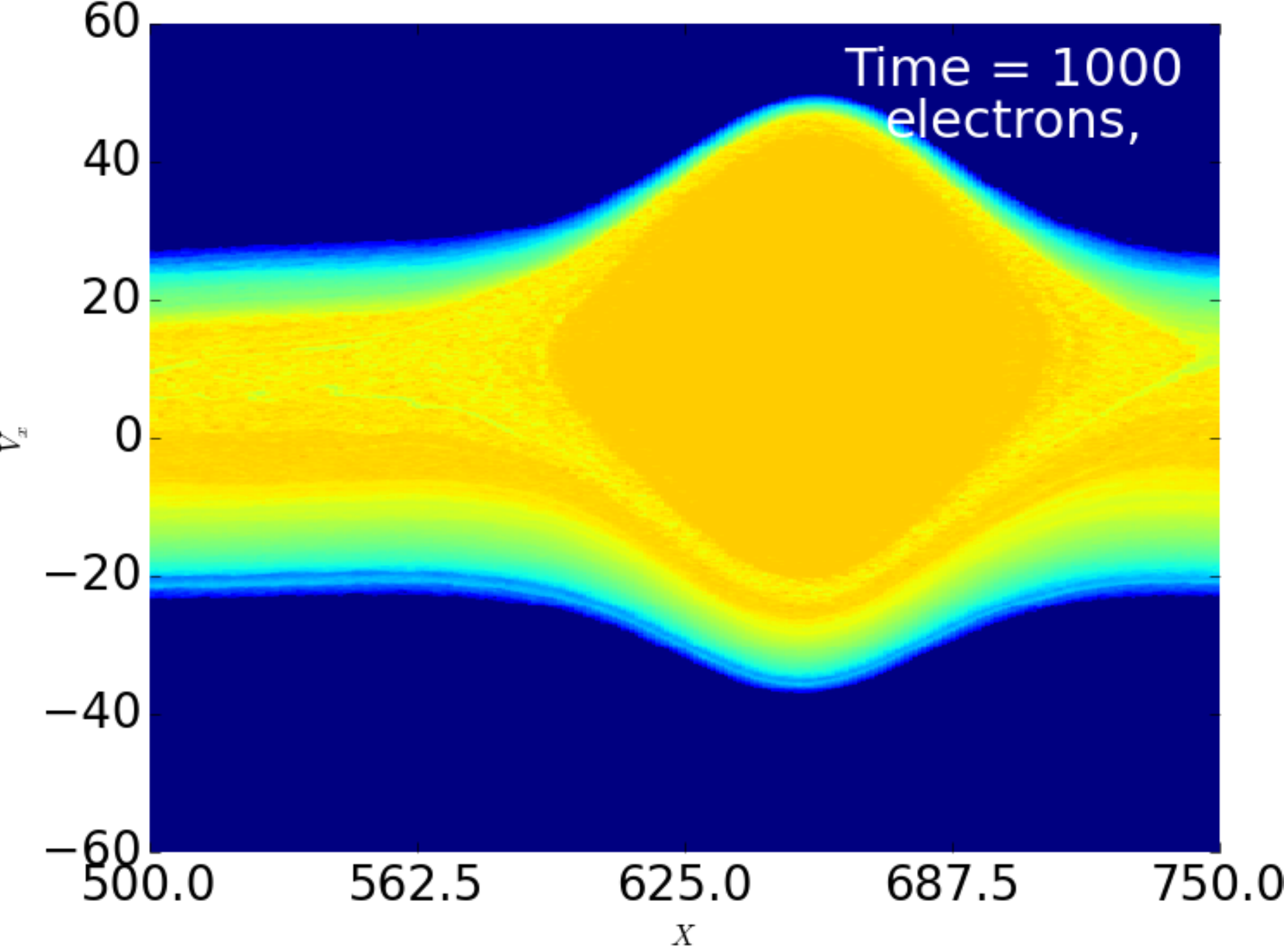}}&  
  \subfloat{\includegraphics[width=0.25\textwidth]{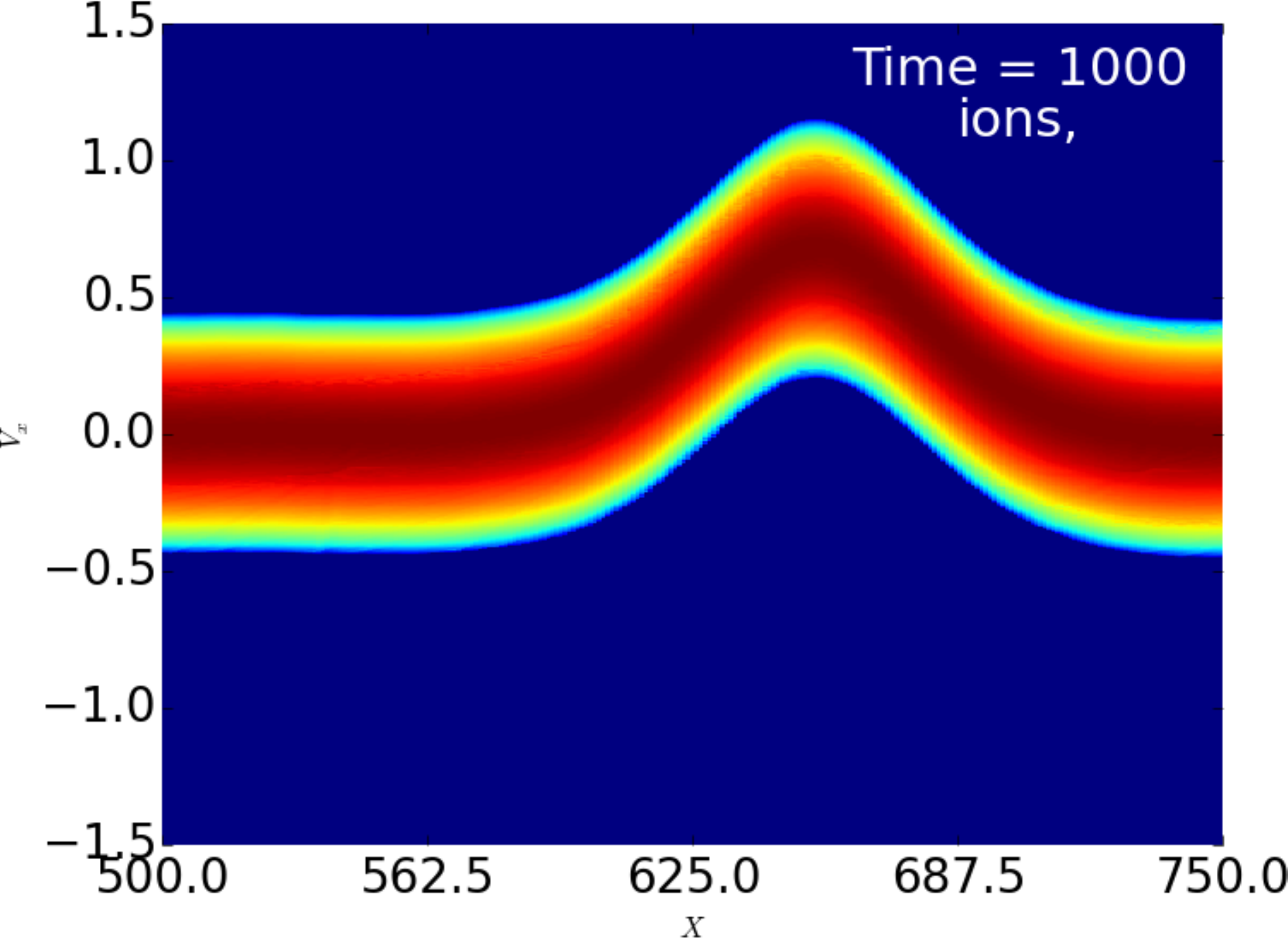}}\\
 \multicolumn{2}{c}{(b) the soliton with $\beta = 0$}\\
 \end{tabular}
  \caption{The phase space structure of the soliton associated 
  to $\beta = 0.2$ and $\beta = 0$ is shown 
  for two different time steps $\tau =0, 1000$.
  This proves the stability of the soliton for 
  a long-time propagation simulation and serve 
  as a benchmarking test of the code..}
  \label{Fig_PlR_ZsR_PhaseSpace}
\end{figure}

As the first step, we have considered the long-time simulation of solitons propagation,
aimed at examining the reliability of the simulation approach utilized here. 
Figs.\ref{Fig_PlR_ZsR_num} and \ref{Fig_PlR_ZsR_PhaseSpace} 
present the results of two solitons with almost the same speed propagating
in a periodically bounded plasma. 
They have different sizes and values of trapping parameter, e.g. $\beta = -0.1, 0.2$.

Fig. \ref{Fig_PlR_ZsR_num} displays the fluid-level physical features, i.e. number densities, for both species. 
Since number density is the starting point for all the other quantities such as electric field and potential, 
their stabilities during propagation guarantee these other quantities stability.
Fig. \ref{Fig_PlR_ZsR_PhaseSpace}, on the other hands, deals with the kinetic details of the distribution function 
of both species during the solitons' long-time propagation. 
The overall shape and internal structure stay the same up-to $\tau = 1000$.
Combining these two figures, all the physical features 
(such as velocity, height and width in both spatial and velocity directions)
of a soliton are preserved by this simulation method
for a long time. 
Considering the ion distribution function around the solitons, 
trapping effect can't be seen and hence the simulations here are in the regime
without ion trapping effect.

The example shown here, provides a benchmarking test for the simulation code. 
Firstly, they prove the the process of inserting solitons distribution functions
into a new simulation box has not resulted in any numerical instability. 
Secondly, it shows that the routines in time step of the simulation code is capable of handling 
a long-time simulation for the nonlinear stage
and justifies the results of the next sections to be reliable.
Note that solitons are extremely sensitive nonlinear structures,
and any numerical errors in the process of 
simulation temporal evolution 
can easily cause them to become unstable and 
deform or disappear from the simulation. 
Finally, this example clearly indicates that the ion-acoustic solitons 
in the presence of the electron trapping can exist
(for more details see Ref. 6). 
The deviation of the total energy and entropy up-to $\tau = 1000$ are $0.2\%$ and $0.05\%$, respectively.

\subsection{Stability of IASs in overtaking collisions } \label{SubSec_Stability}
\begin{figure}
  \subfloat{\includegraphics[width=0.5\textwidth]{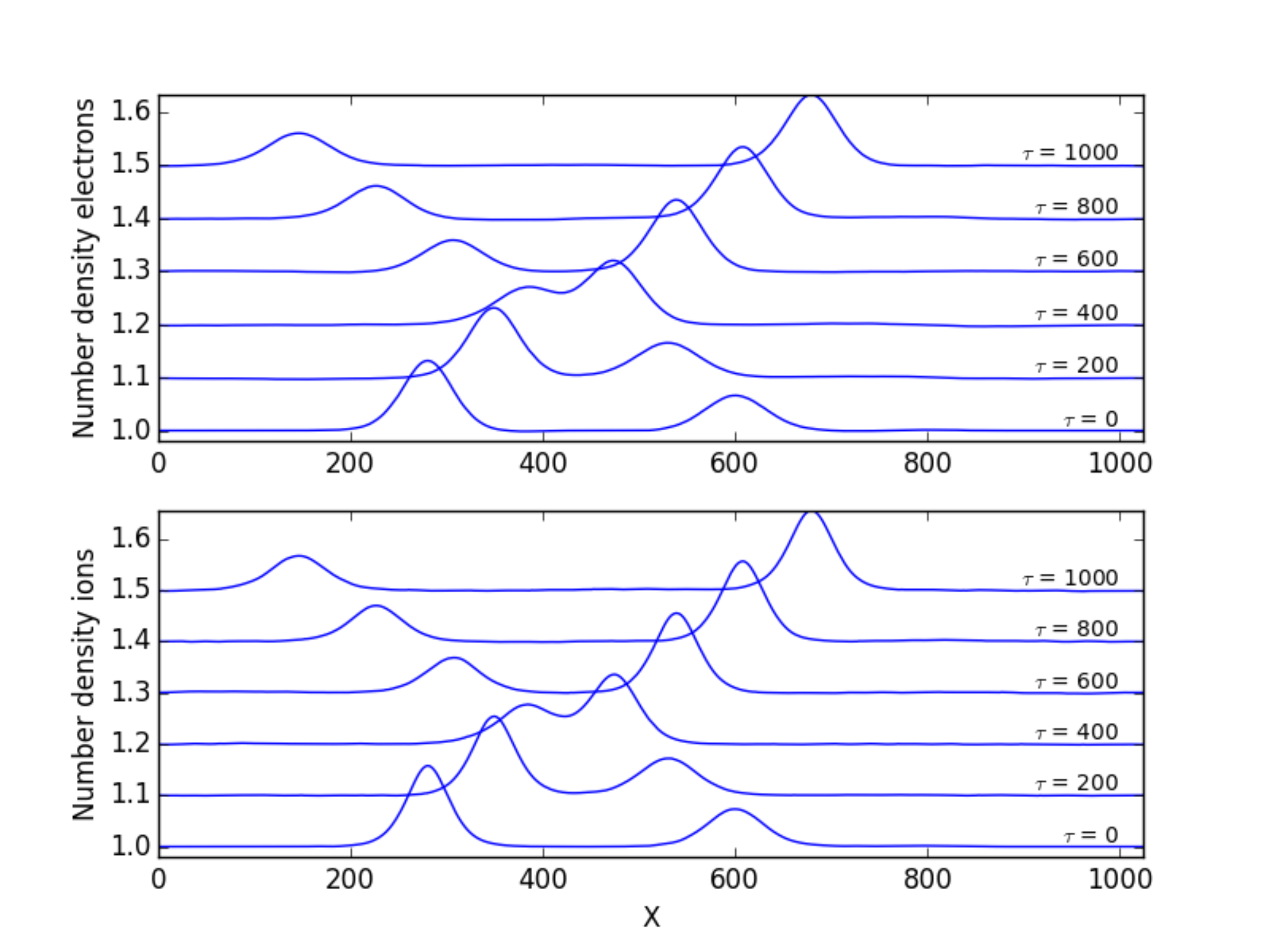}}
  \caption{Overtaking collision is shown for two solitons with same $\beta = -0.1$ by sketching 
  the number densities of electrons and ion in the window moving with their average speed($v = 9.45$).
  Overtaking happens around $\tau = 400$, and results are shown until $\tau = 1000$ to display the stability of 
  solitons after the overtaking.}
  \label{Fig_NlR_NsR_num}
\end{figure}

\begin{figure}
 \begin{tabular}{c c}
  \subfloat{\includegraphics[width=0.26\textwidth]{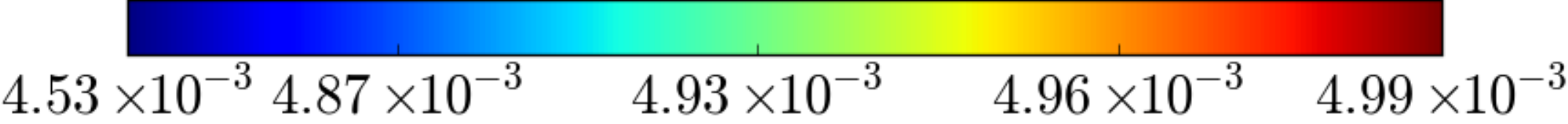}}&
  \subfloat{\includegraphics[width=0.26\textwidth]{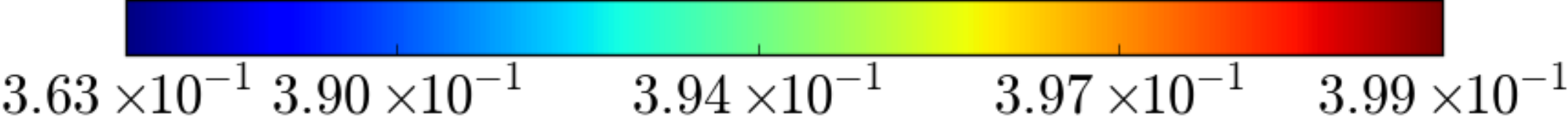}} \\
  \subfloat{\includegraphics[width=0.25\textwidth]{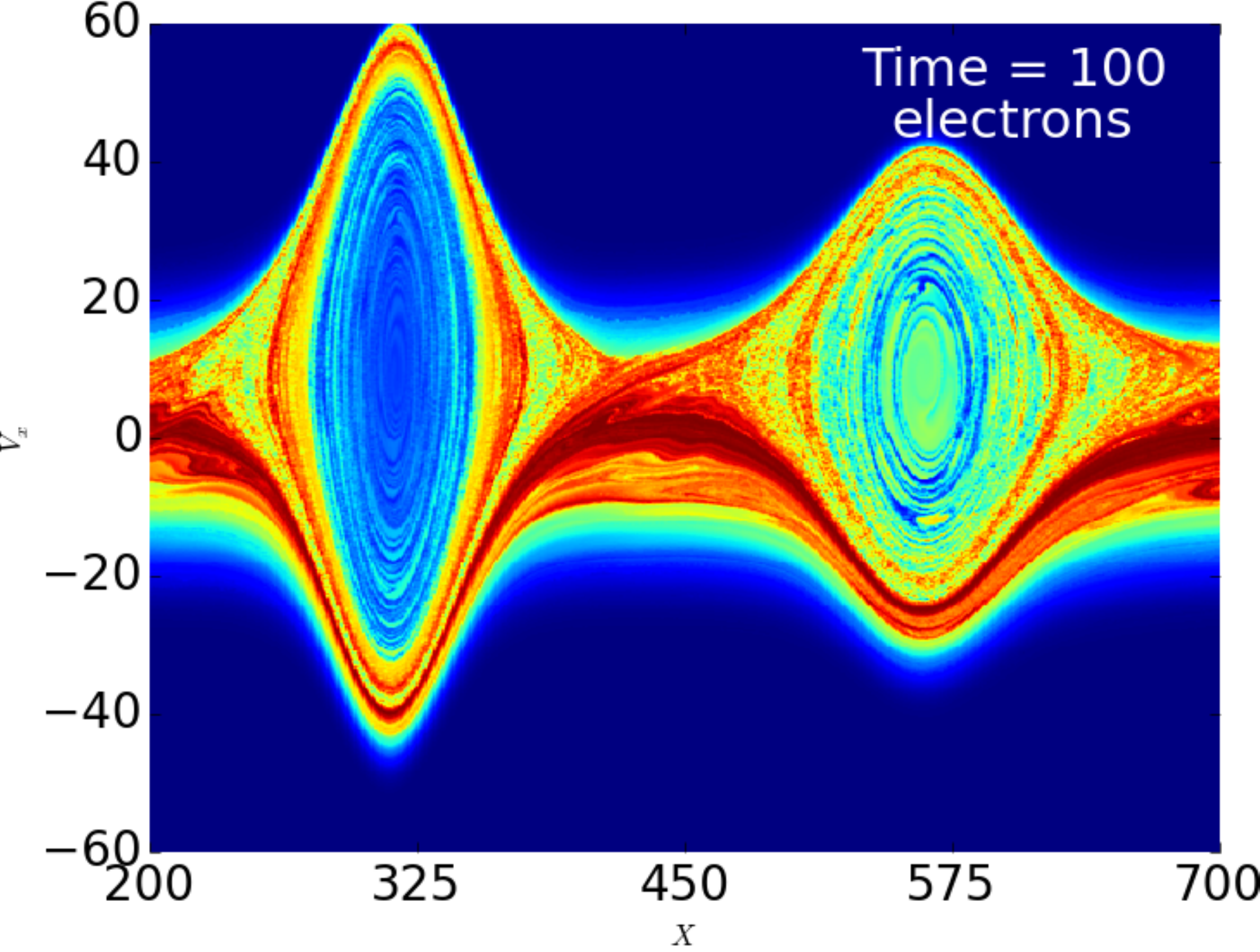}}&
  \subfloat{\includegraphics[width=0.25\textwidth]{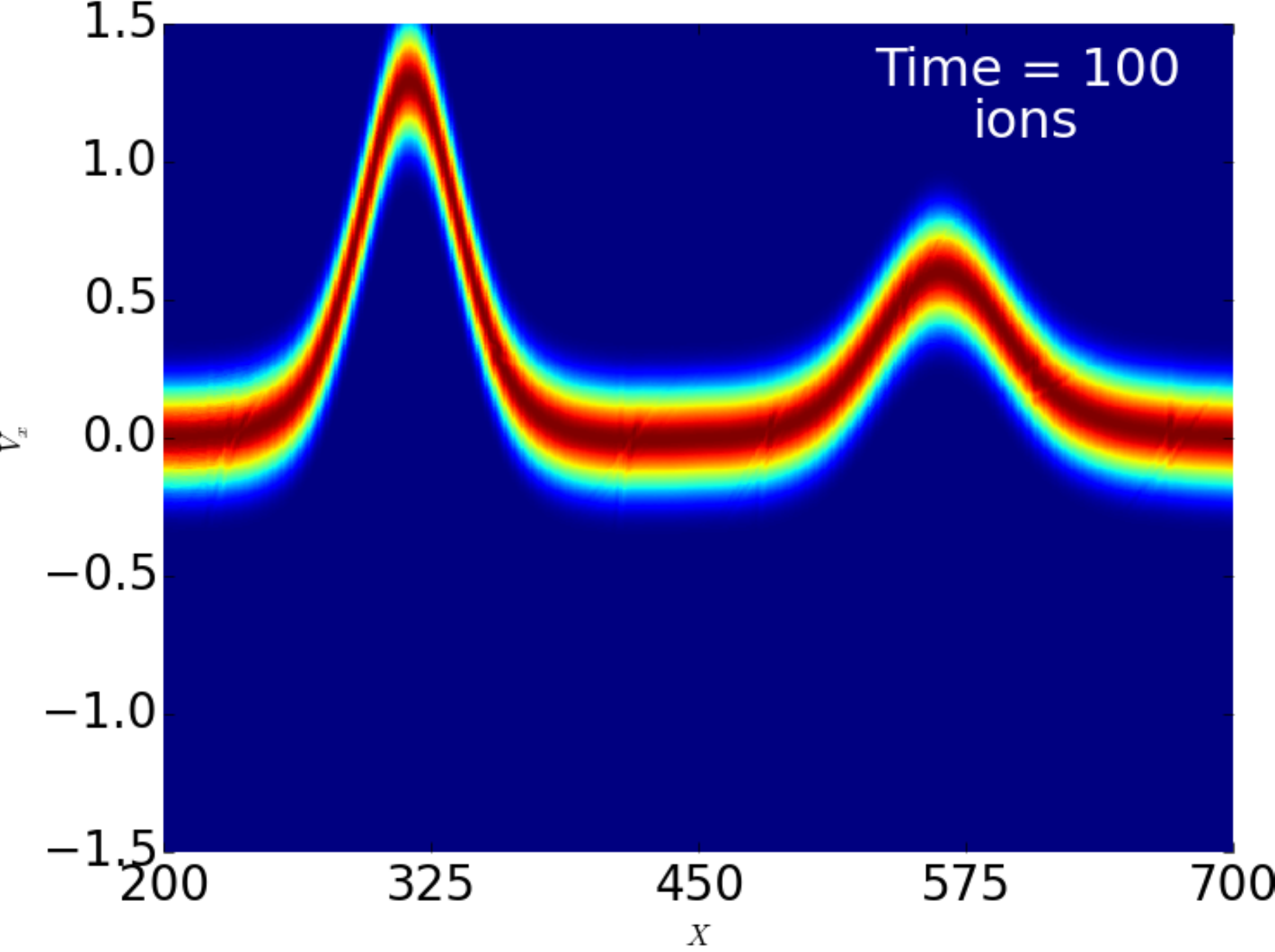}} \\
  \subfloat{\includegraphics[width=0.25\textwidth]{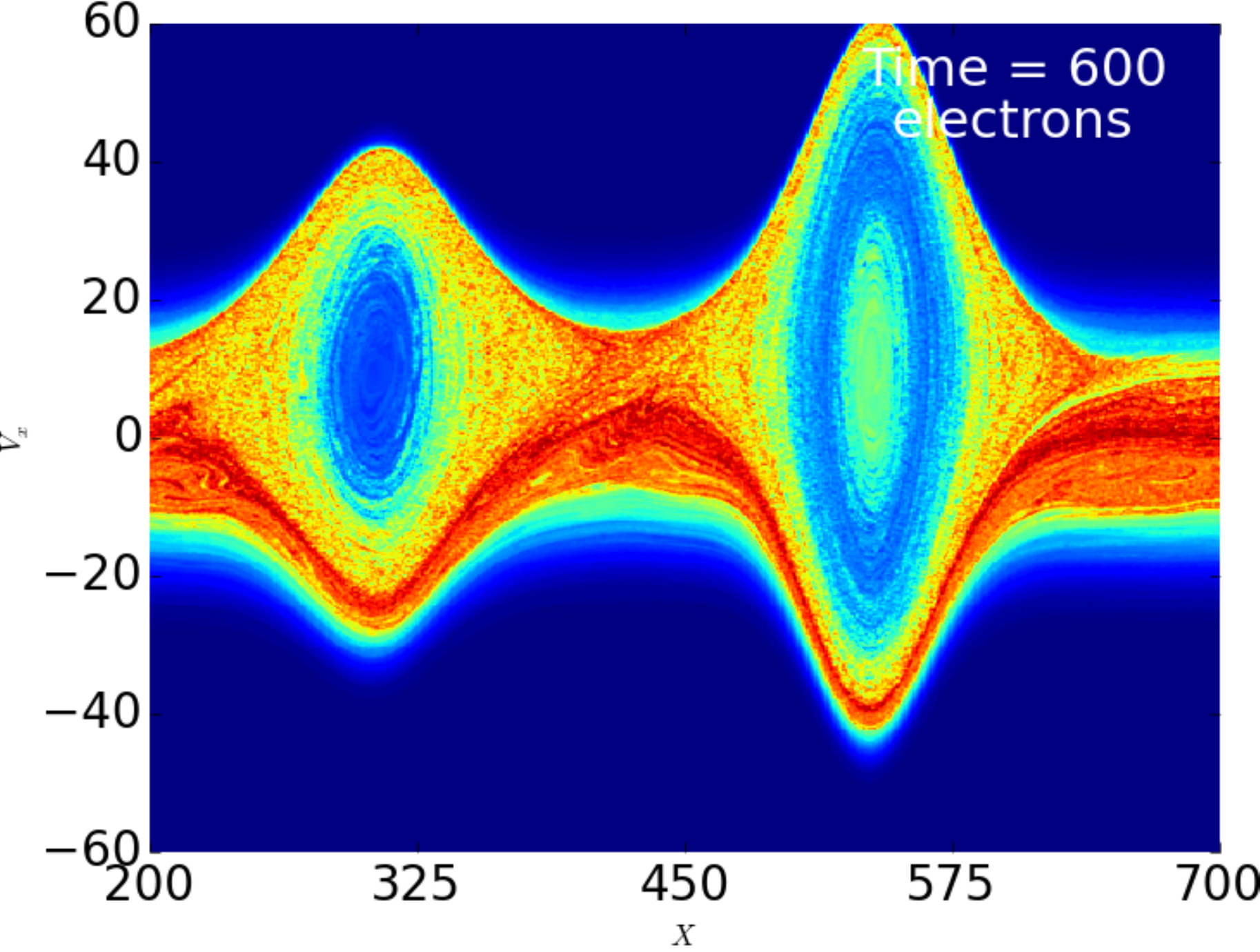}}&  
  \subfloat{\includegraphics[width=0.25\textwidth]{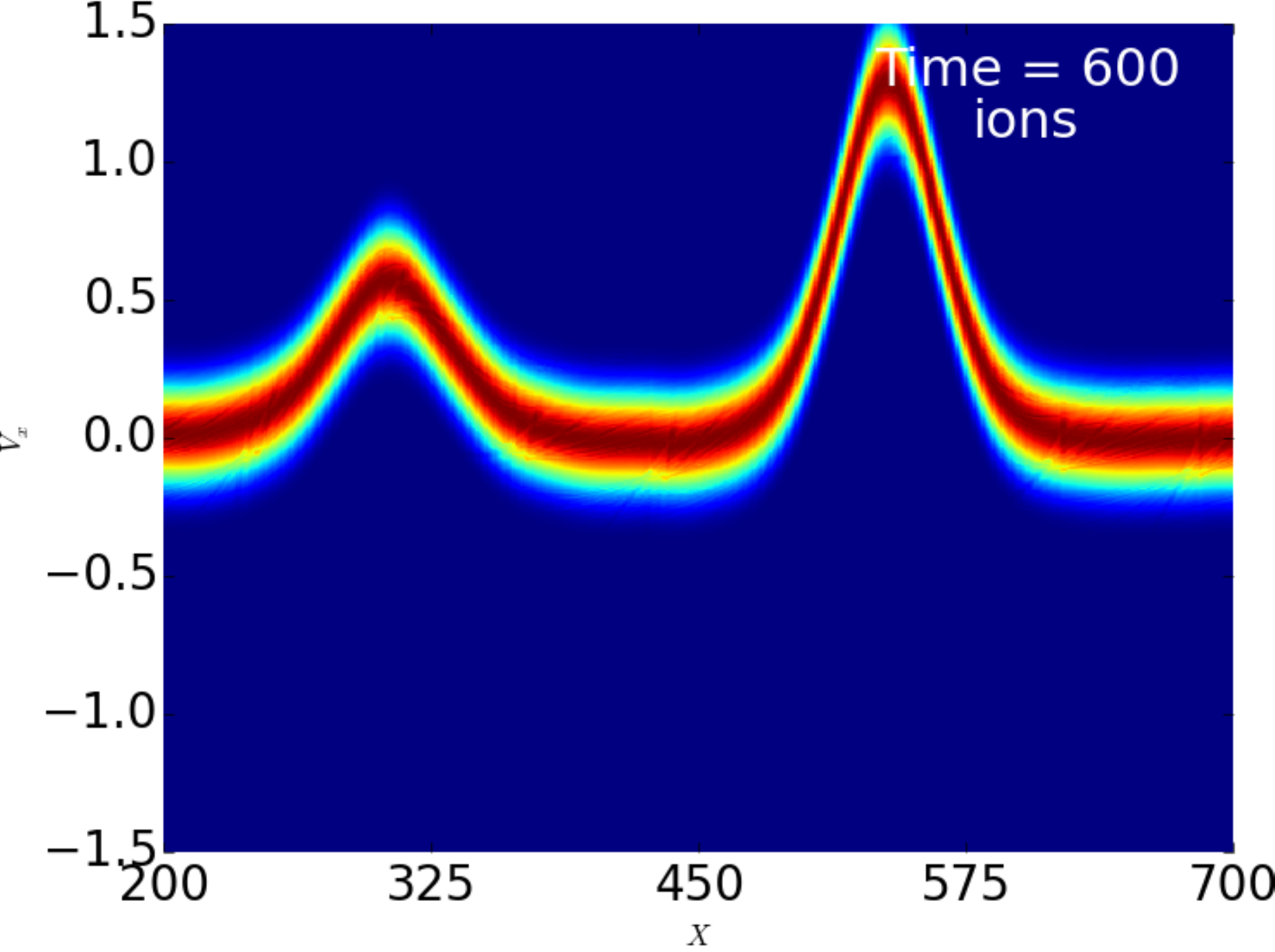}} 
 \end{tabular}
  \caption{Phase space of ions and electrons are shown for before ($\tau = 100$) and 
  after ($\tau = 600$) the first overtaking collision between.
  Solitons acquire the same shape in the electrons distribution function, i.e. holes ($\beta = -0.1$).
  They are shown in the frame moving with their average velocity ($v=9.45$)
  The two IASs exchange some parts of their trapped electron population.}
  \label{Fig_NlR_NsR_PhaseSpace}
\end{figure}

Two cases of overtaking collisions between solitons are presented here.
In the first case (Figs. \ref{Fig_NlR_NsR_num} and \ref{Fig_NlR_NsR_PhaseSpace})
IASs with same trapping parameters $\beta = -0.1$
collide while the larger soliton overtaking the smaller one.
Second case is dedicated to solitons with different trapping parameter 
($\beta = -0.1$ and $\beta = 0$). 
The results prove the stability of IASs in presence of trapping effect of electrons
against mutual overtaking collisions.

In case of $\beta = -0.1$, the overtaking collision of IASs happens 
while they are accompanied by two holes in electron distribution function. 
Fig. \ref{Fig_NlR_NsR_num} shows the number densities, fluid-level features, before and after the collision. 
On the fluid level, the stability against mutual collision can be witnessed, 
since solitons' features such as hight, shape and width remain the same before and after the collision. 
Note that the collision is defined as the time interval of solitons when they are overlapping each other. 
Hence the times $\tau < 200$ and $\tau > 600$ are considered as before and after the collision, respectively.
During collision/overlapping (for example at time $\tau = 400$ shown in Fig. \ref{Fig_NlR_NsR_num}), 
the two solitons lose their distinctive shape and merge to some extent. 
However, the concept of stability is defined for solitons features 
before and after collisions and not during them.

Furthermore, the kinetic details of both species distribution functions before and after collision
is shown in Fig. \ref{Fig_NlR_NsR_PhaseSpace}. 
Although the overall shape, width in velocity and spatial directions remain the same,
the internal structures differ.
The contrast of colors clearly indicates that the after the collision, 
each of the electron holes has acquired some of 
the trapped electrons' population of the opposite electron hole. 
Each trapped population can be recognized after the collision by their core (inner part).
In other words, the core remains untouched during collision.
However, the outer (parts) are exchanged between the two electron holes.
In case of ions distribution function no change can be seen before and after the collision.

\begin{figure}
  \subfloat{\includegraphics[width=0.5\textwidth]{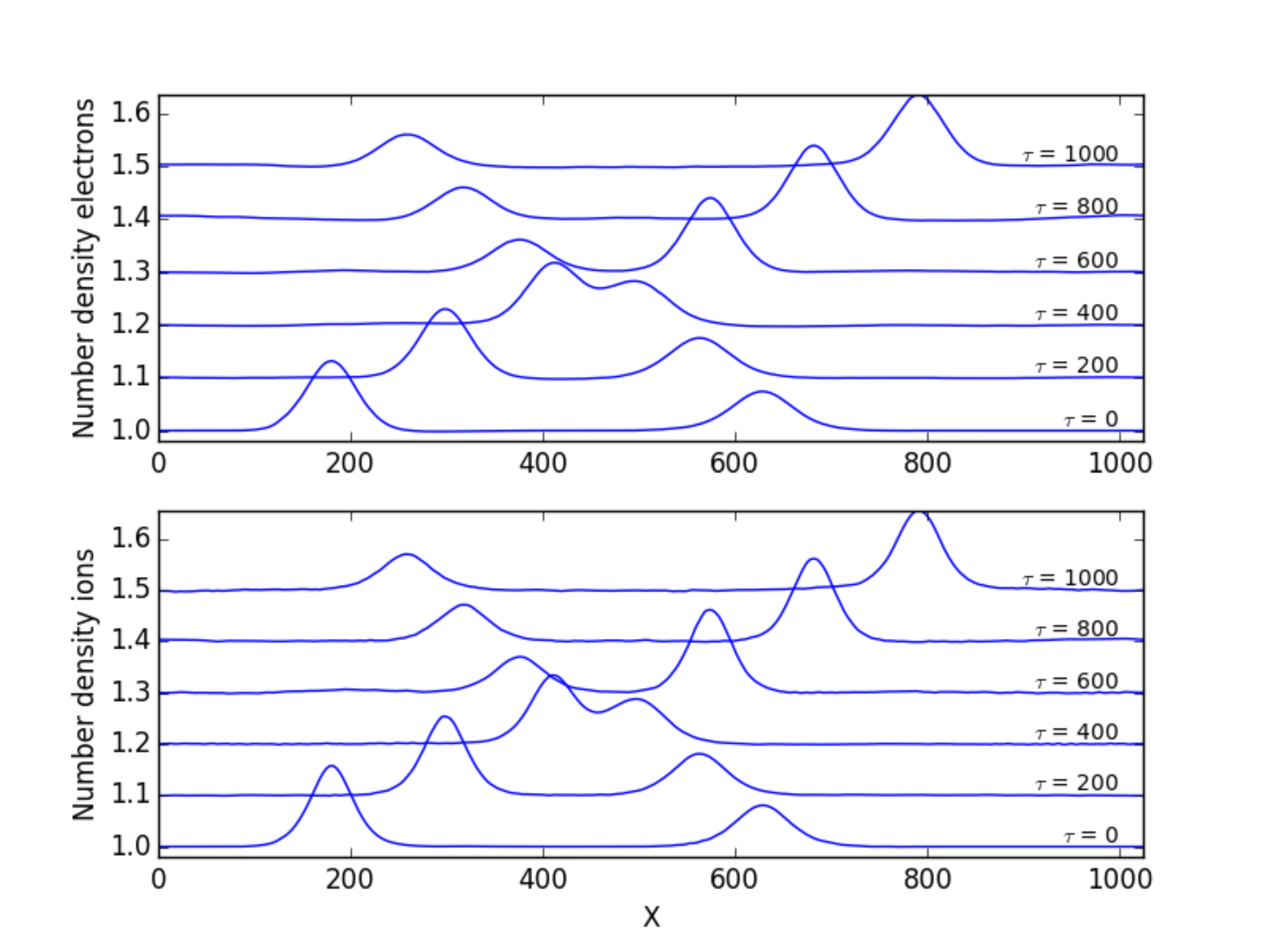}}
  \caption{Overtaking collision is shown for two solitons with different trapping parameter, 
  i.e. $\beta =-0.1$ and $\beta = 0$, by sketching 
  the number densities of electrons and ion in the frame moving with their average speed($v = 9.8$).
  Overtaking happens around $\tau = 400$, and results are shown until $\tau = 1000$ to display the stability of 
  solitons after the overtaking.}
  \label{Fig_NlR_ZsR_num}
\end{figure}

\begin{figure}
  \begin{tabular}{c c}
    \subfloat{\includegraphics[width=0.23\textwidth]{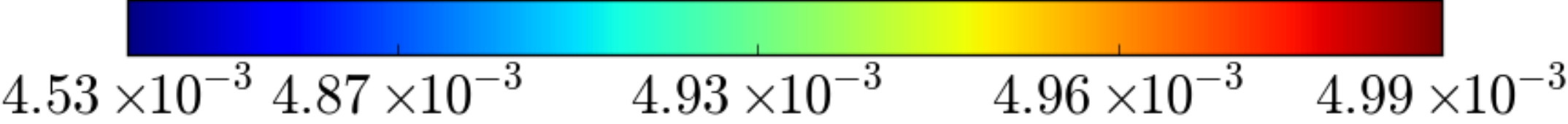}}&
    \subfloat{\includegraphics[width=0.23\textwidth]{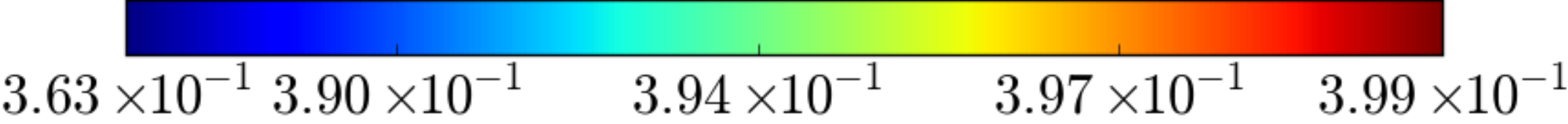}} \\
    \subfloat{\includegraphics[width=0.25\textwidth]{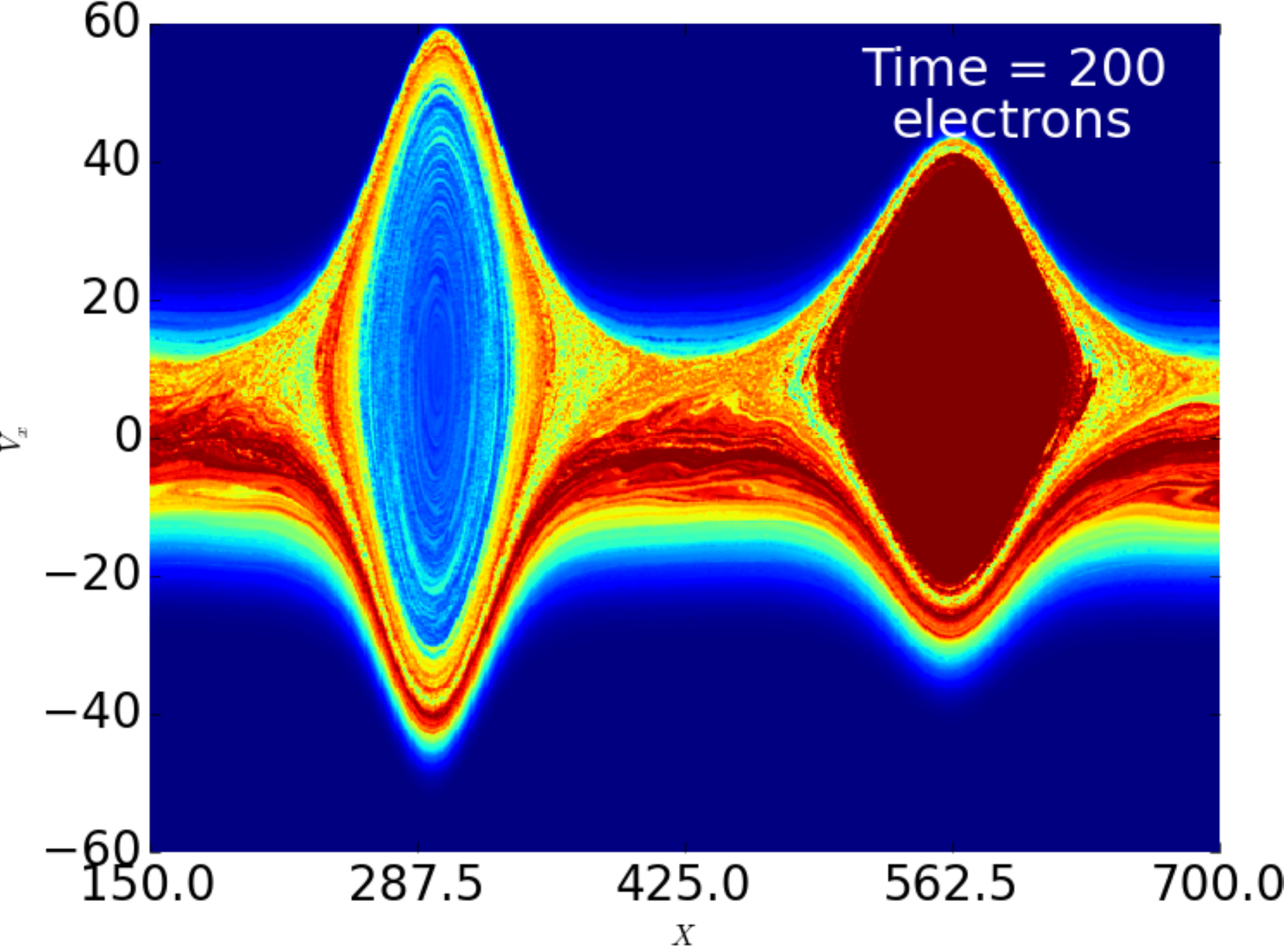}}&
    \subfloat{\includegraphics[width=0.25\textwidth]{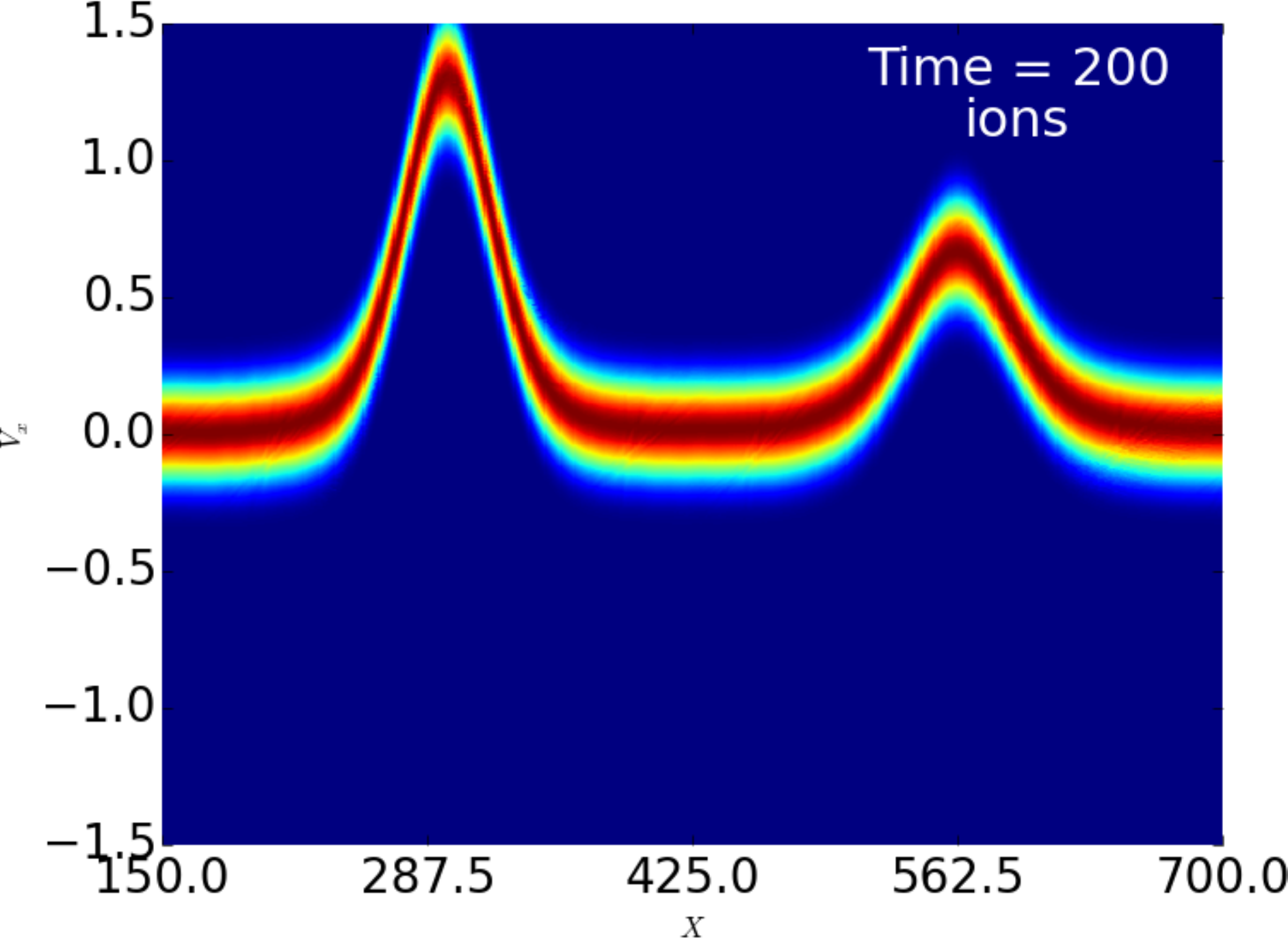}} \\
    \subfloat{\includegraphics[width=0.25\textwidth]{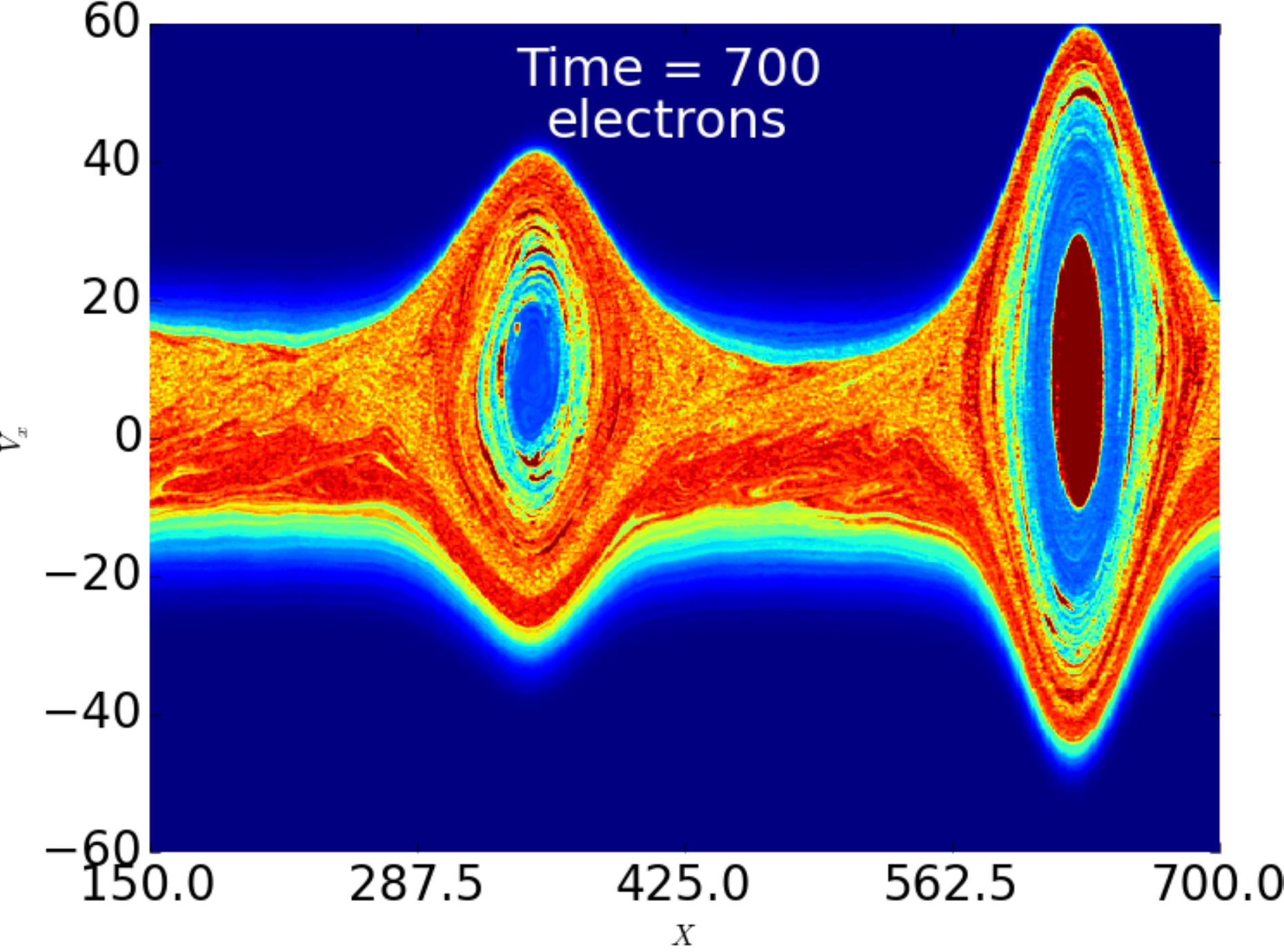}} &  
    \subfloat{\includegraphics[width=0.25\textwidth]{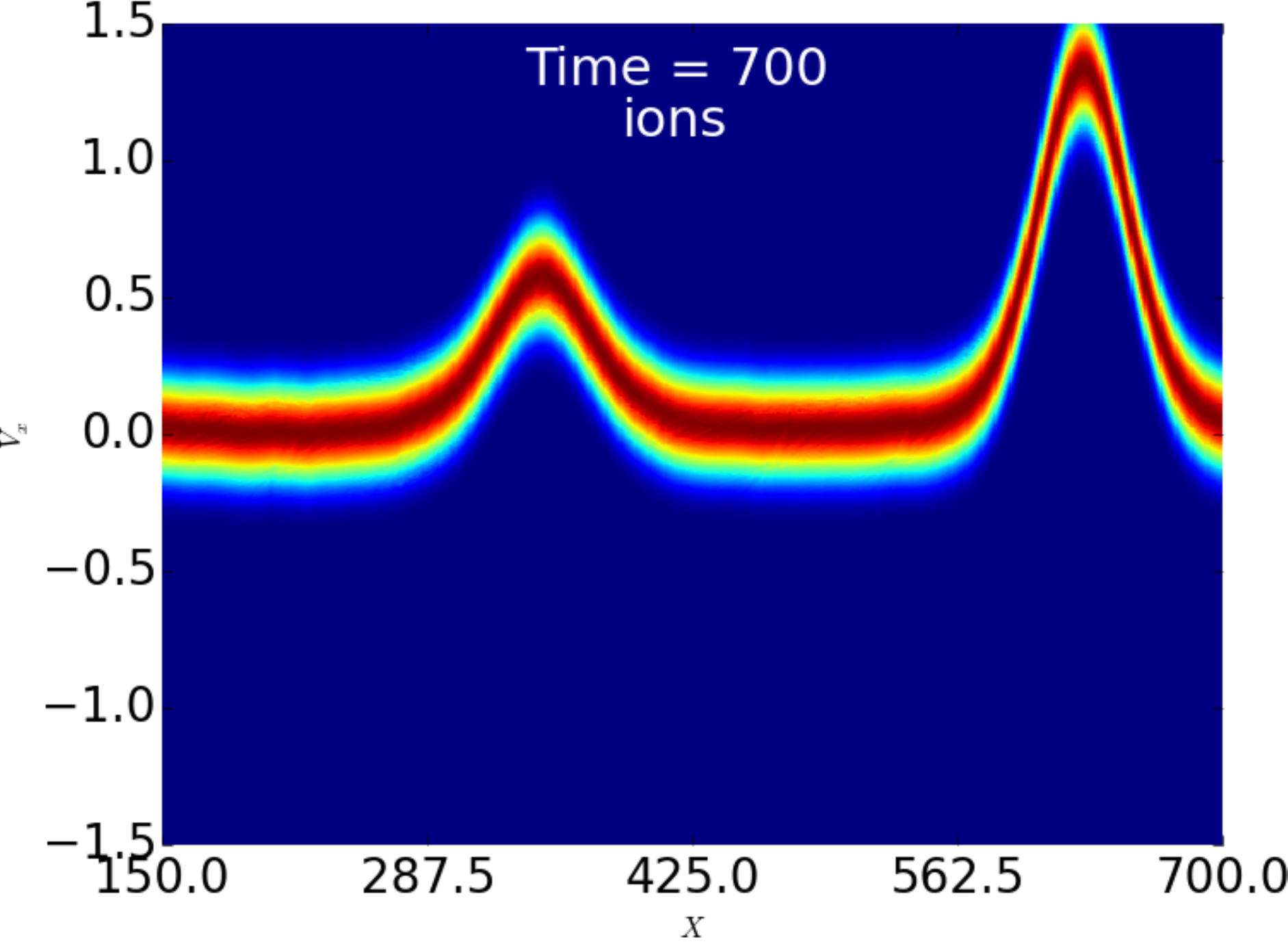}} 
  \end{tabular}  
  \caption{Phase space of both species, i.e. electrons and ions, are show for two solitons with different value of
  trapping parameters, $\beta = -0.1$ (hole) and $\beta = 0$ (plateau).
  Two snapshots associated to before ($\tau = 100$) and after ($\tau = 600$) are presented
  in the frame moving with their average velocity $v=9.8$.
  The trapped population of electrons interchange their outer parts
  with each other while their cores stay the same.
  }
  \label{Fig_NlR_ZsR_PhaseSpace}
\end{figure}

\begin{figure}
  \subfloat[$\beta = -0.1$]{\includegraphics[width=0.25\textwidth]{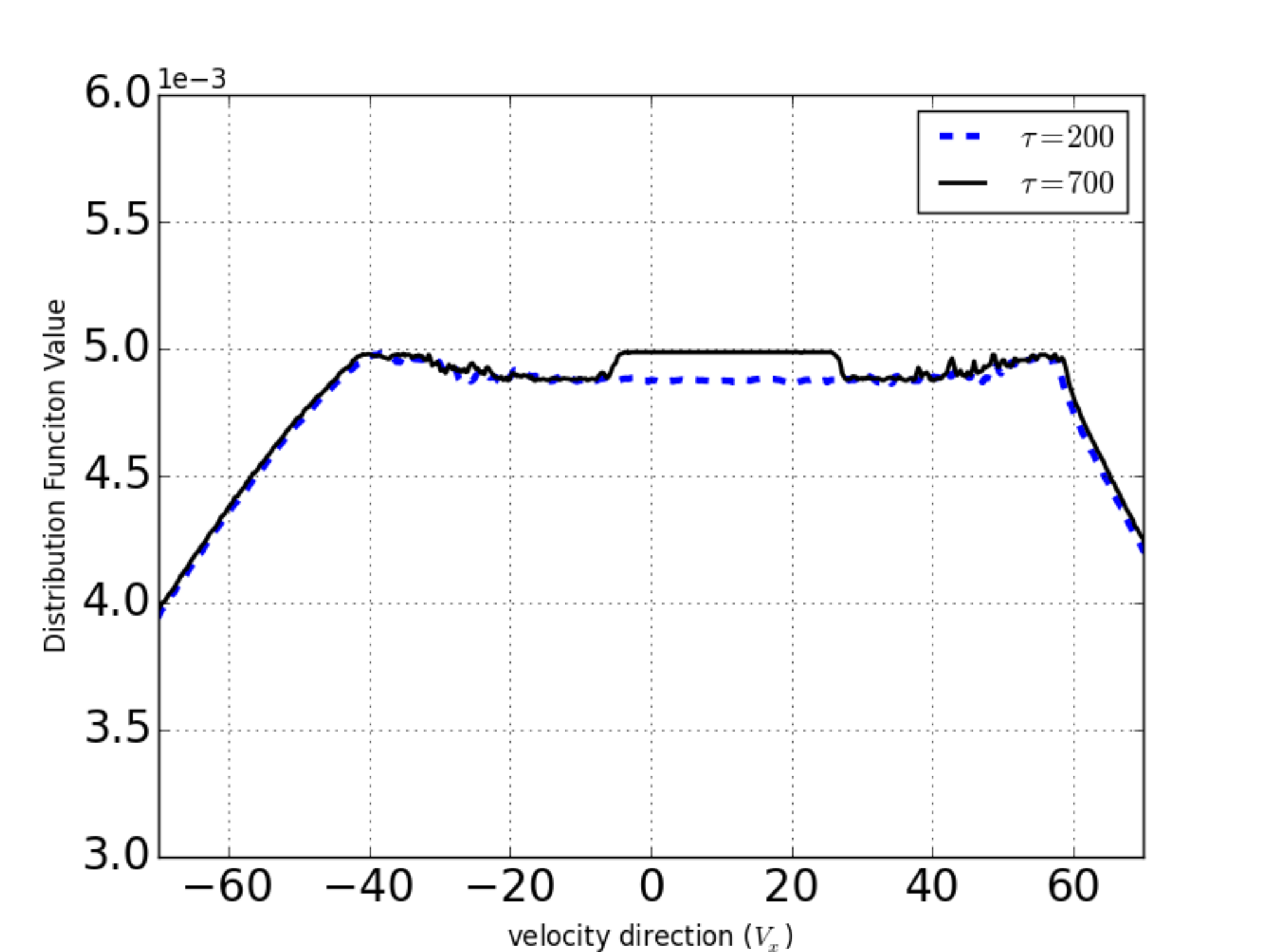}}
  \subfloat[$\beta = 0.0$]{\includegraphics[width=0.25\textwidth]{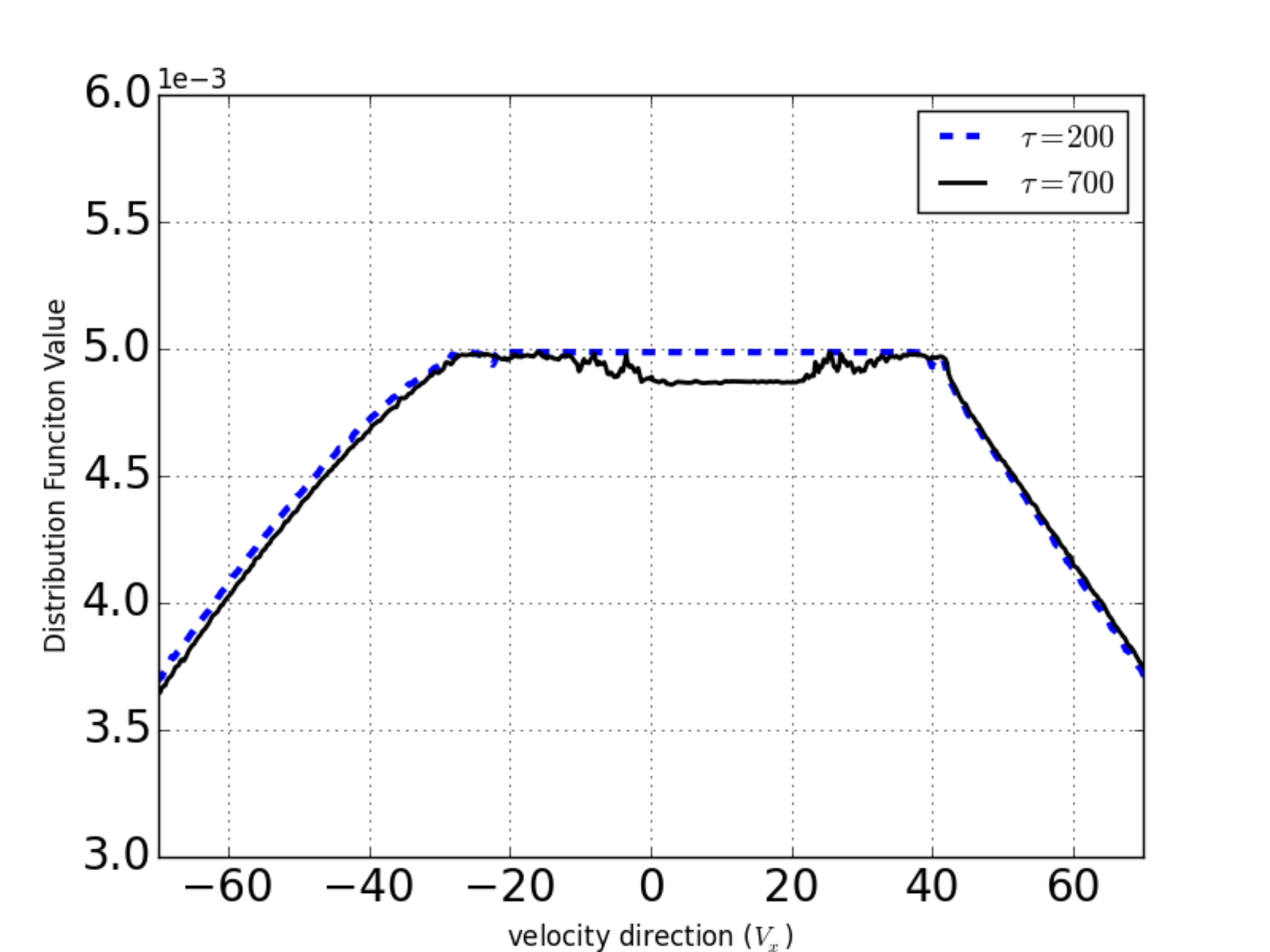}}
  \caption{Electrons distribution functions for (a) $\beta = -0.1$ and (b)$\beta = 0.0$ are shown before ($\tau = 200$) and after ($\tau = 700$) 
  overtaking collision between the two solitons.
  The exchange of population around $V_x = 0$ can be witnessed, as the soliton with negative trapping parameter $\beta = -0.1$ (with a hole shape in phase space)
  has acquired \textbf{some parts of trapped population} from $\beta = 0$ (with a plateau shape).}
  \label{Fig_NlR_ZsR_cross}
\end{figure}

Figs.\ref{Fig_NlR_ZsR_num} and \ref{Fig_NlR_ZsR_PhaseSpace}
display the results of an overtaking collision between 
two solitons with different trapping parameter, namely $\beta = 0$ and $\beta = -0.1$. 
Hence on the kinetic level of electron, the collision takes place
between a plateau ($\beta = 0$) and a hole ($\beta = -0.1$)
accompanying small and large solitons, respectively. 
The same characteristics as the collision between two electron holes can be witnessed 
such as exchanging outer layers of trapped populations and 
no change in ion distribution function.
\textbf{Fig.\ref{Fig_NlR_ZsR_cross} clearly indicates the exchange of population between the two solitons.
Furthermore it implies the conservation of the trapped partilces. }

\subsection{Details during overtaking collisions} \label{SubSec_During}
\begin{figure}
  \subfloat{\includegraphics[width=0.6\textwidth]{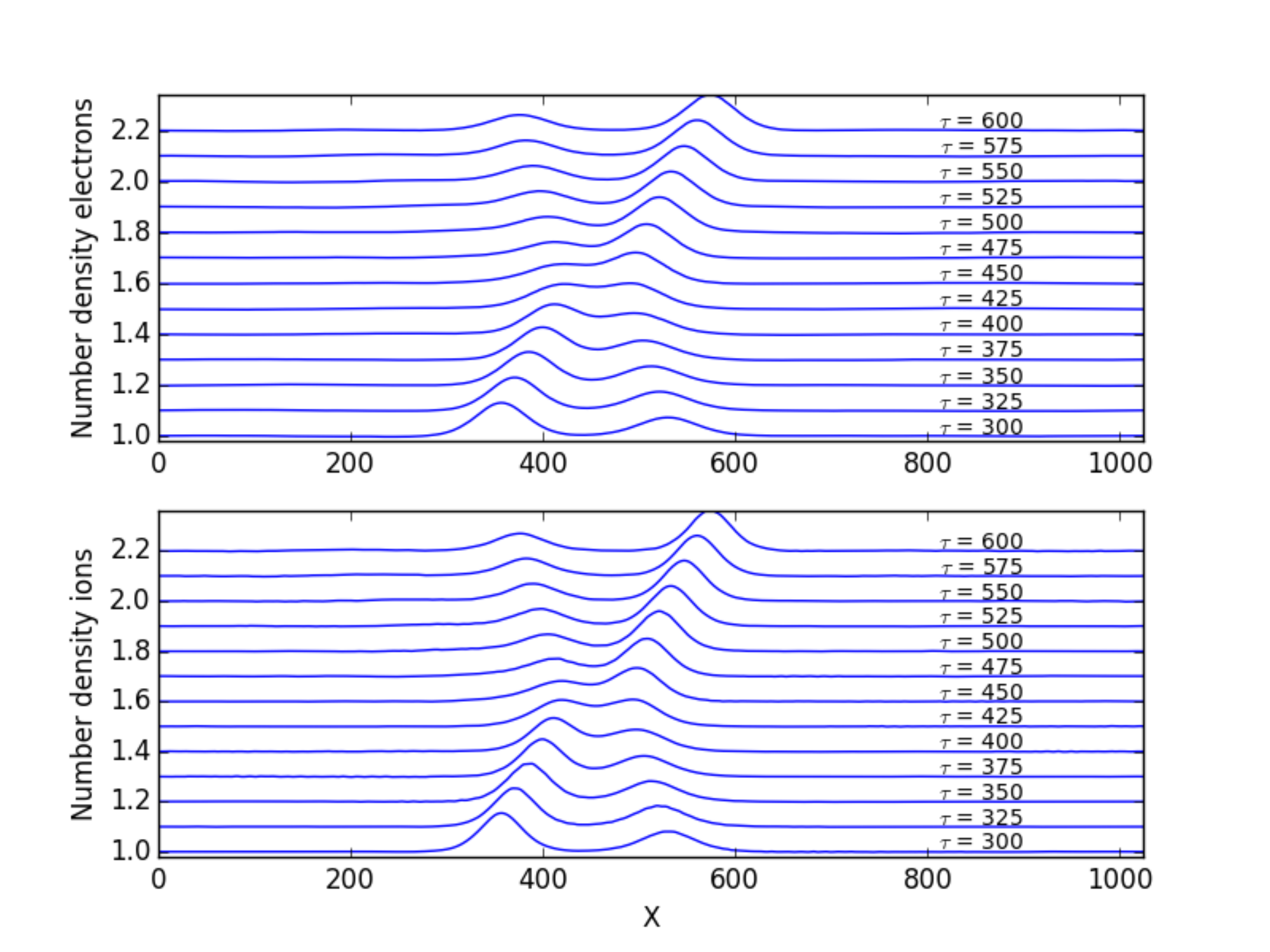}}
  \caption{Temporal evolutions of ions and electrons number densities are plotted
  for solitons with different different trapping parameters, 
  i.e. $\beta = -0.1$ (hole) and $\beta = 0$ (plateau), during their overtaking collision.
  Phase shift can be seen in the trajectories of both IASs, as before $\tau<425$ and after collision, 
  the trajectory of large/small IASs is shifted to right/left.
  Note that the plots are sketched in a moving frame with the average velocity
  of the solitons.}
  \label{Fig_NlR_ZsR_num_details}
\end{figure}

Fig. \ref{Fig_NlR_ZsR_num_details} displays the 
details of temporal evolution of number densities of the two species during
the overtaking collision. 
Three interesting phenomena can be witnessed in this figure.

Firstly, Fig. \ref{Fig_NlR_ZsR_num_details} reveals that 
during the overtaking collision, the solitons don't cross each other. 
The larger (faster) soliton reaches the smaller (slower) soliton, 
while losing its height, hence slowing down. 
Meanwhile, the smaller (slower) soliton increase its height and therefore its velocity. 
Note that there is a direct relationship between velocity and height of an IAS.
As the time of $\tau = 425$, the two soliton have the same height and velocity. 
Afterwards, the process of losing/gaining height and velocity by larger/smaller soliton continues. 
Until the smaller/larger one morphs itself completely to the opposite soliton.
Further on, due to the velocity difference they start to depart each other, however this time 
the larger soliton appears ahead of the smaller one. 

Secondly, a shift in the trajectories of both solitons can be recognized 
which is a specific property of collision between solitons and have been reported in context of 
other field of physics for variety of solitons. 
This shift can be conceived as the by product of the two solitons exchanging their features during collision. 
The larger soliton disappears during collision and emerges later in the place of the smaller one
and hence it doesn't follow a continuous path. 
This bending of trajectory appears as a shift seen in Fig. \ref{Fig_NlR_ZsR_num_details}.

Finally, during collision, two solitons overlap and their facing tales 
disappears. 
This overlapping can be seen at $\tau = 425$ in Fig. \ref{Fig_NlR_ZsR_num_details} 
at the midst of the collision. 

\begin{figure}
\begin{tabular}{c c}
  \subfloat{\includegraphics[width=0.25\textwidth]{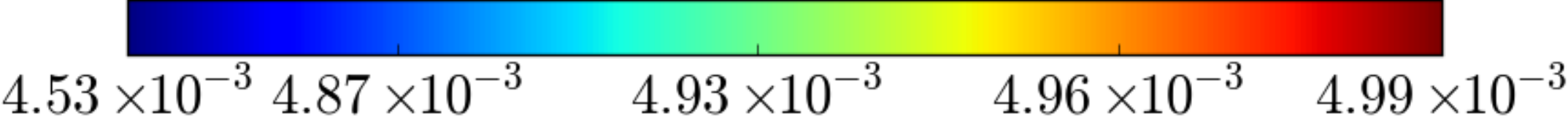}}&
  \subfloat{\includegraphics[width=0.25\textwidth]{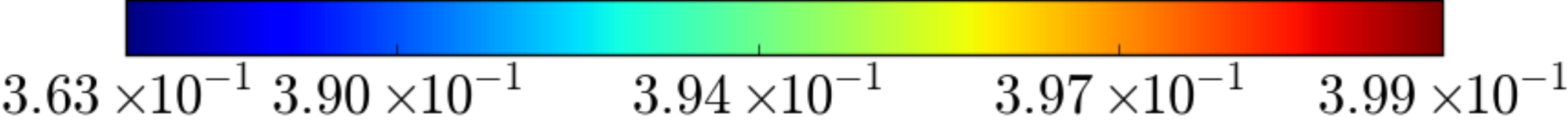}} \\

  \subfloat{\includegraphics[width=0.25\textwidth]{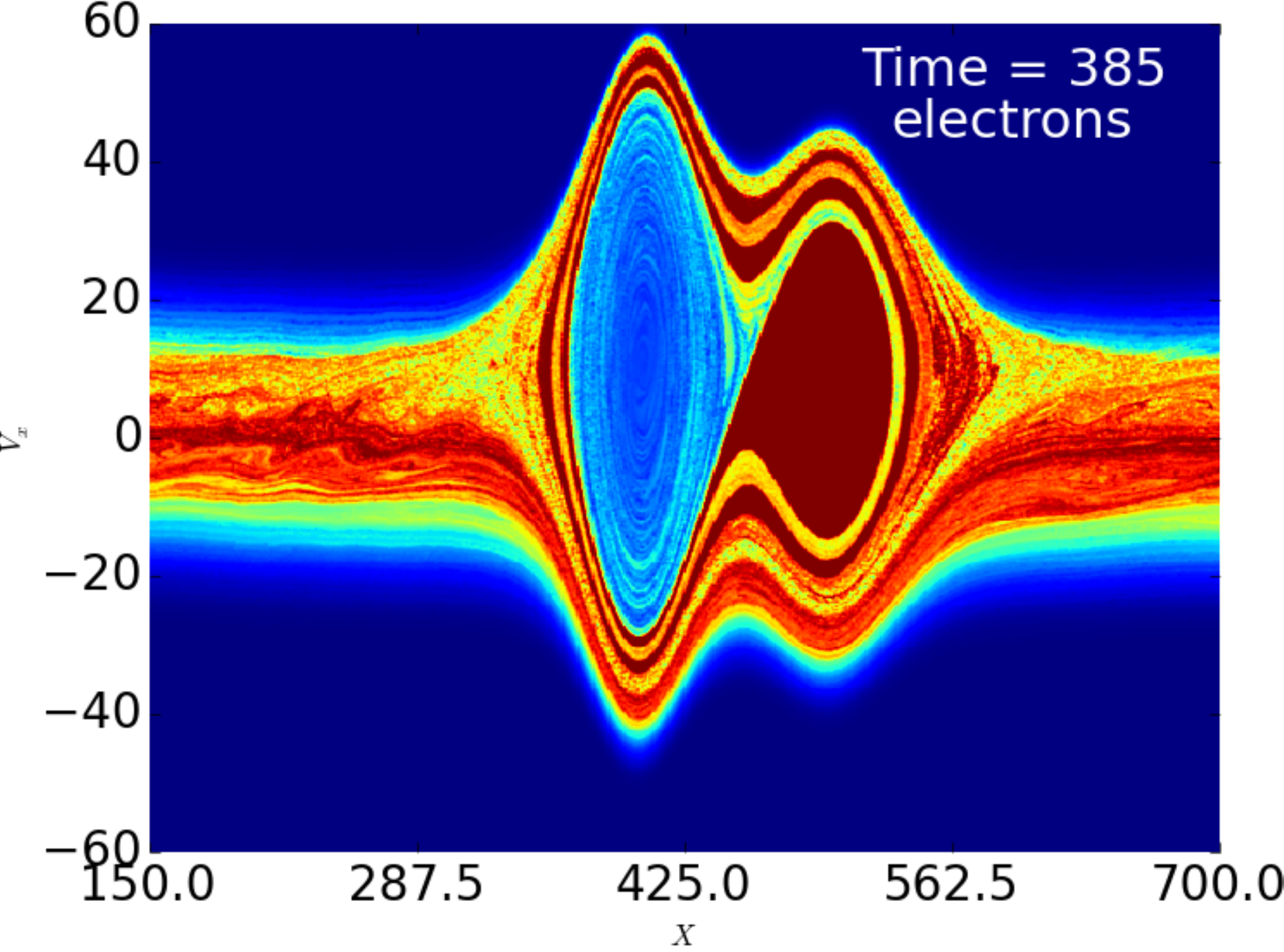}}&
  \subfloat{\includegraphics[width=0.25\textwidth]{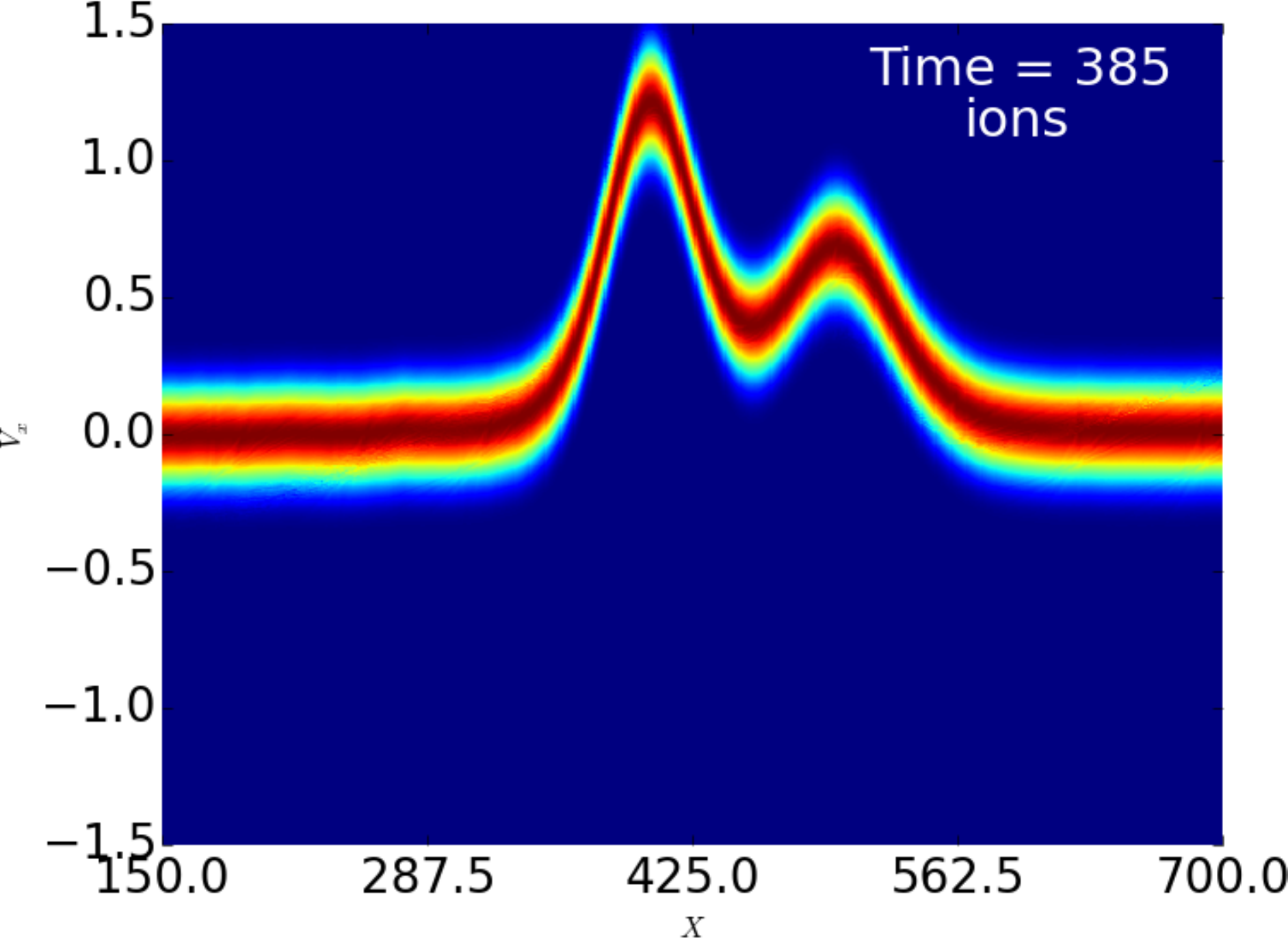}}\\
  
  \subfloat{\includegraphics[width=0.25\textwidth]{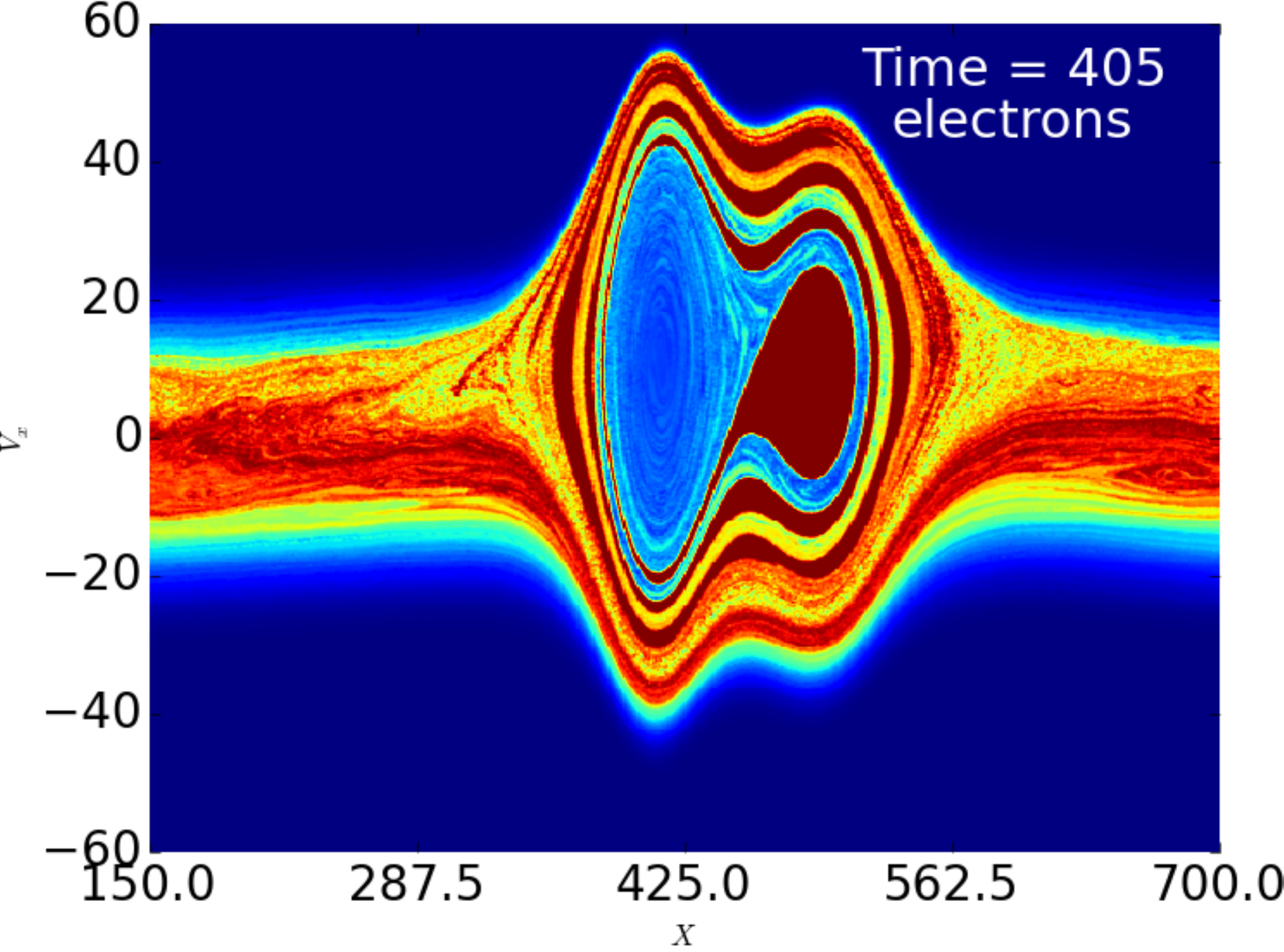}}&
  \subfloat{\includegraphics[width=0.25\textwidth]{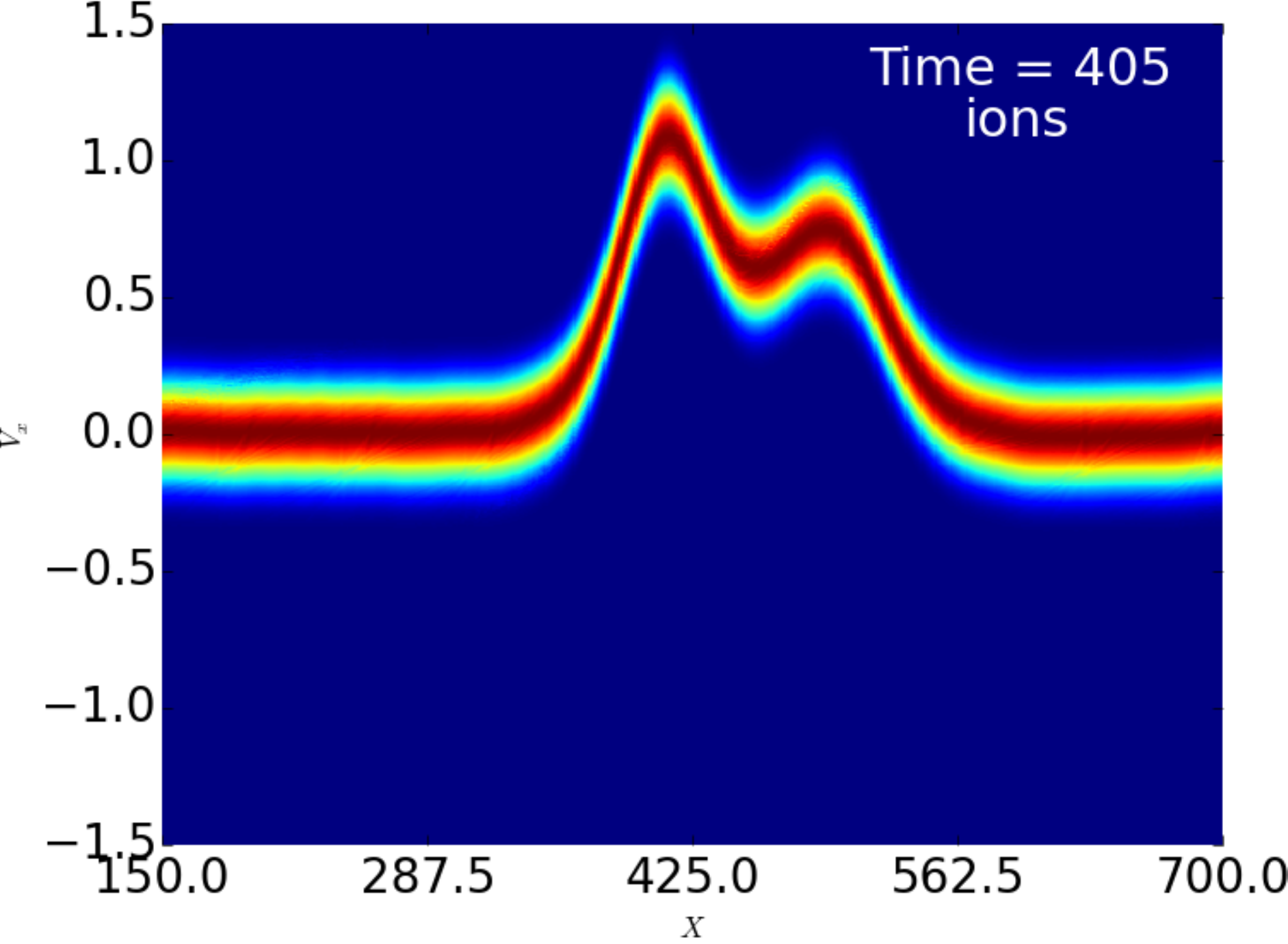}}\\
  
  \subfloat{\includegraphics[width=0.25\textwidth]{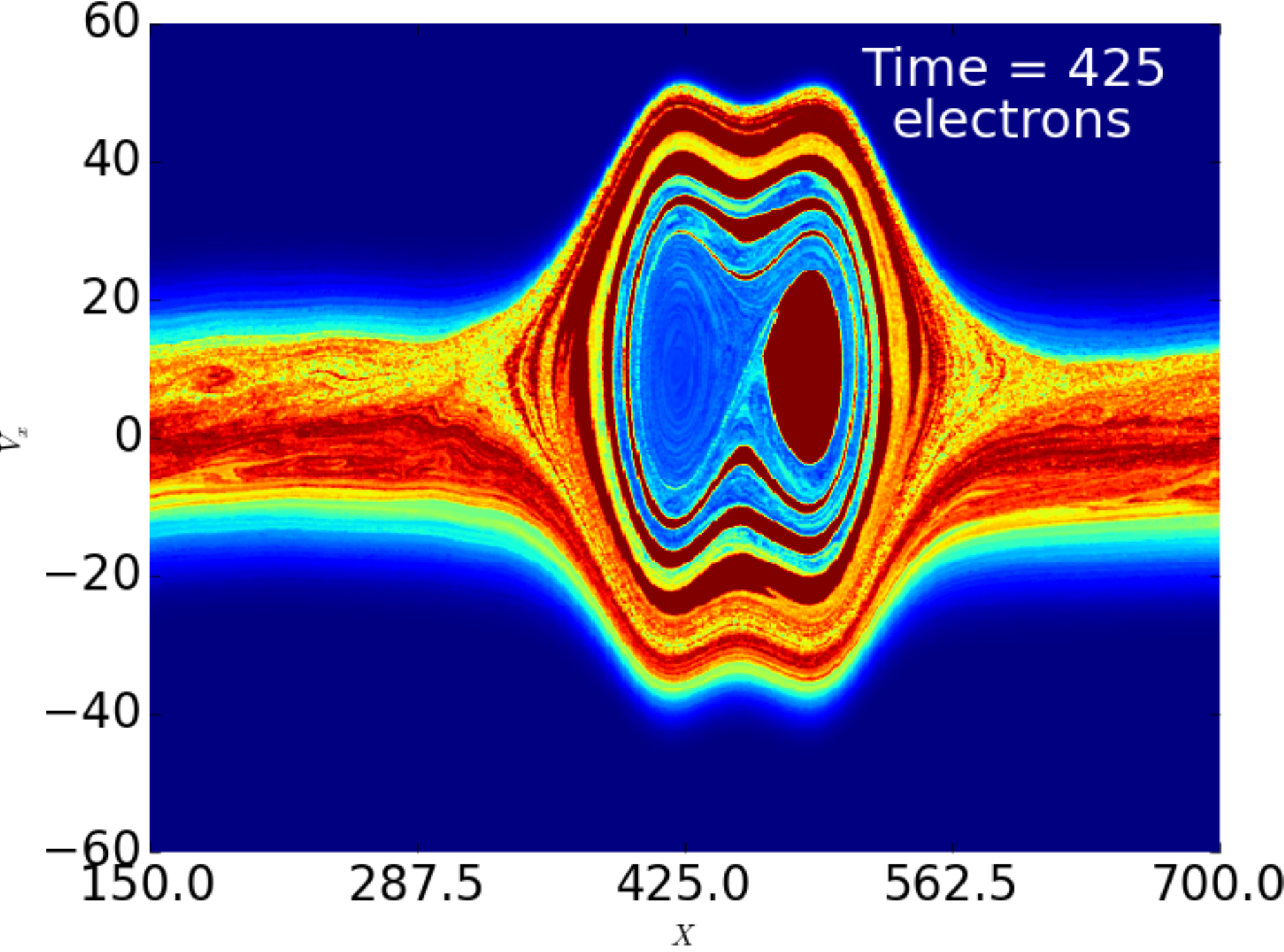}}&
  \subfloat{\includegraphics[width=0.25\textwidth]{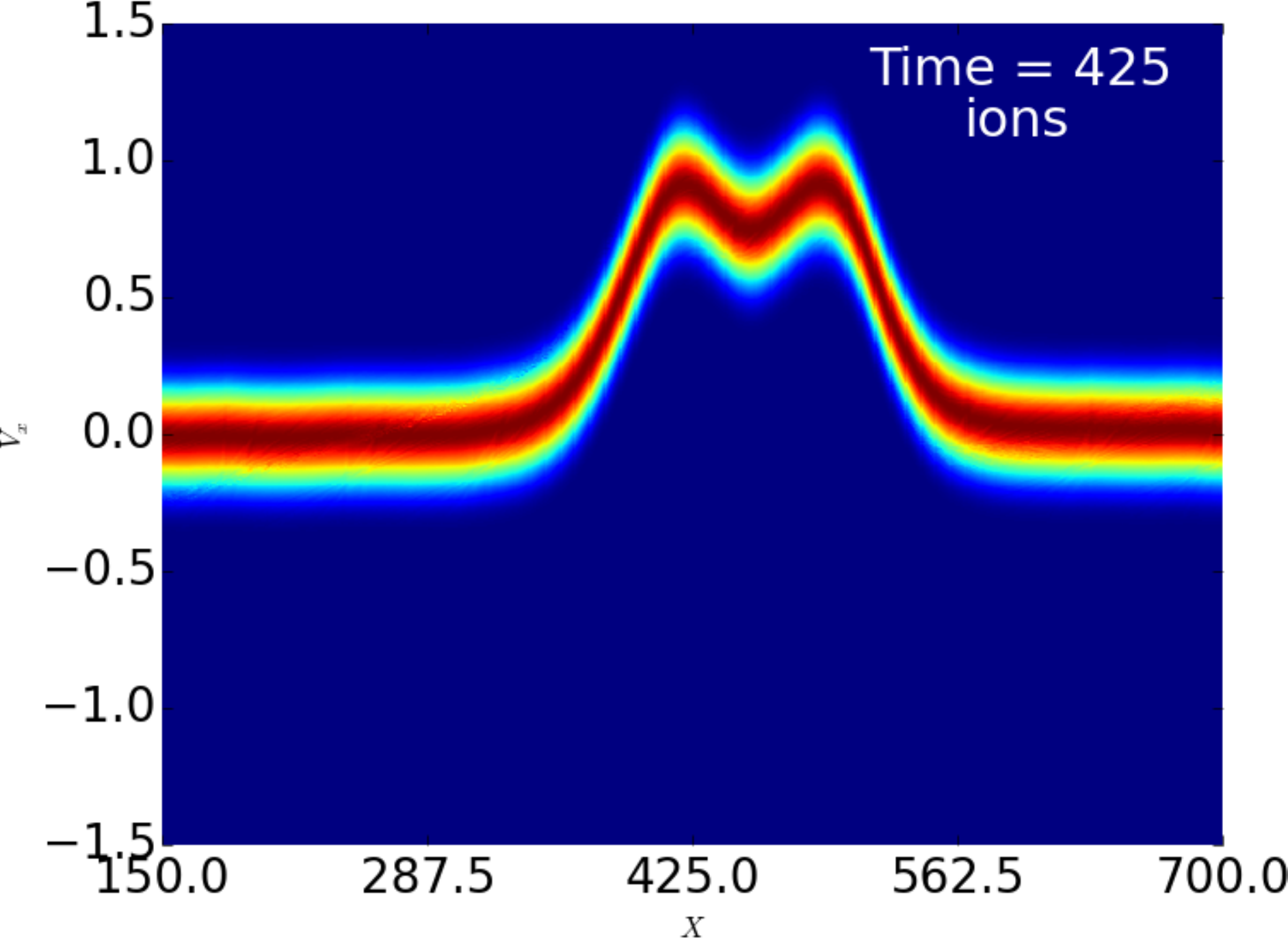}}\\
  
  \subfloat{\includegraphics[width=0.25\textwidth]{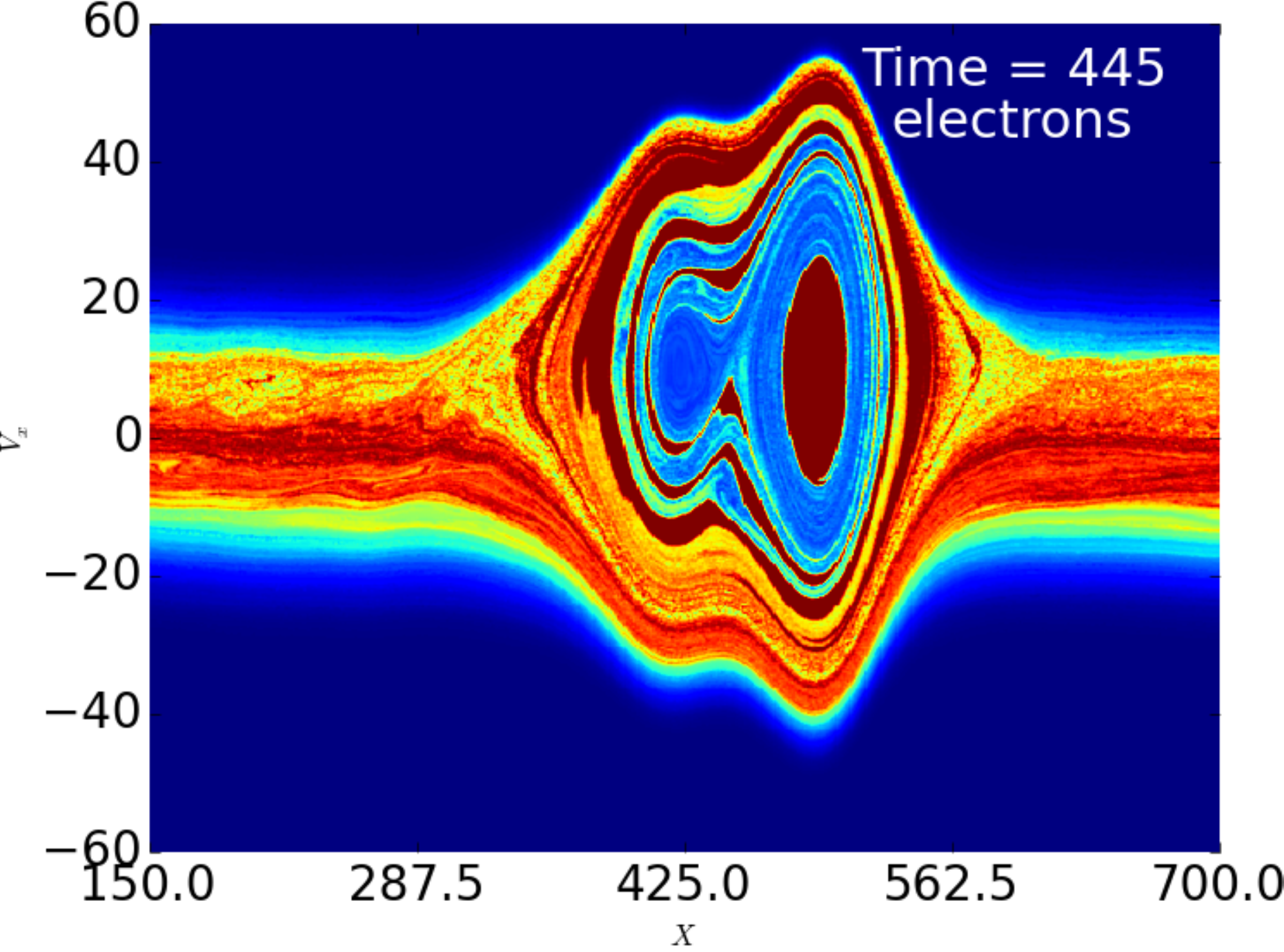}}&
  \subfloat{\includegraphics[width=0.25\textwidth]{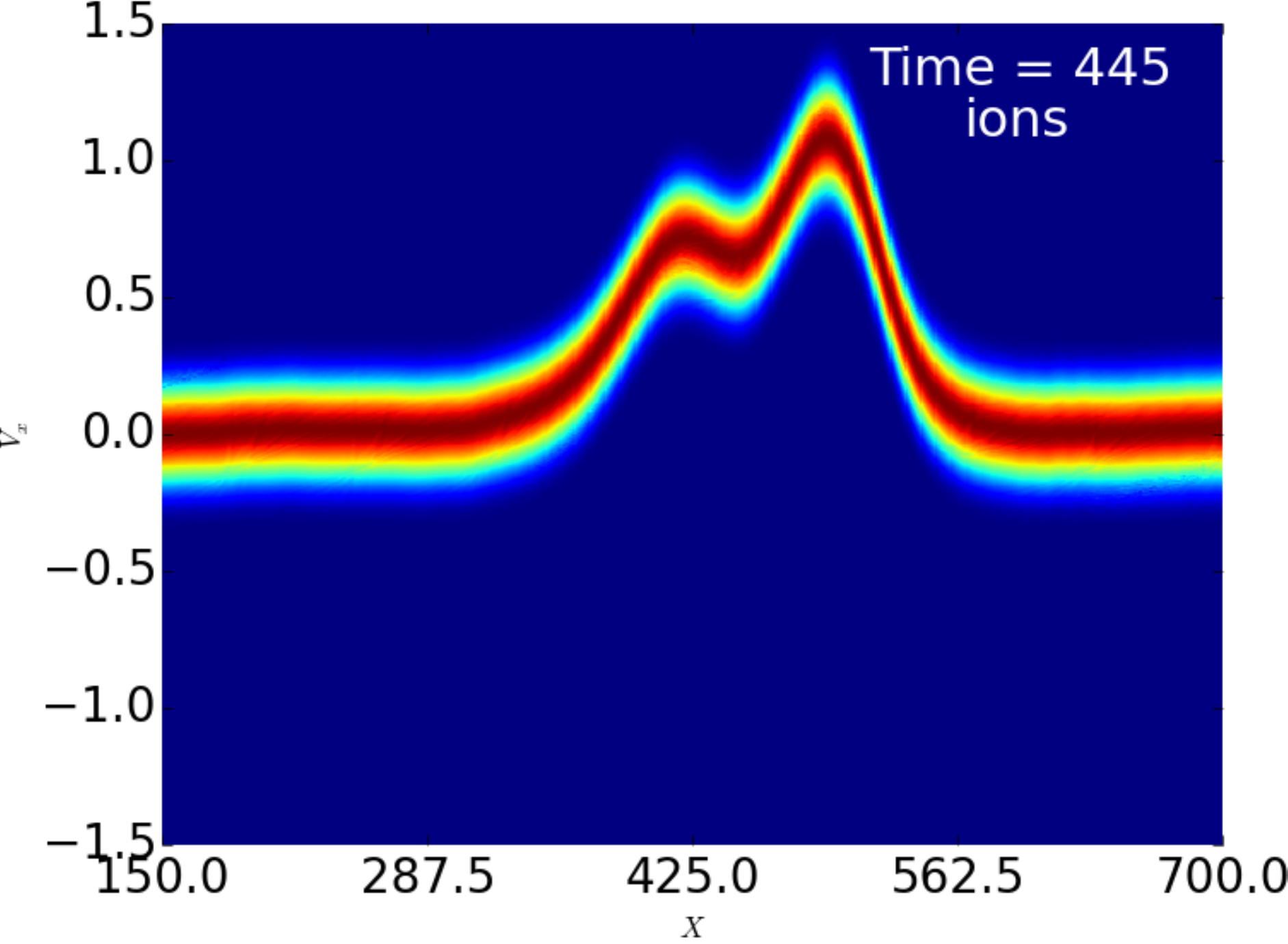}}\\
  
  \subfloat{\includegraphics[width=0.25\textwidth]{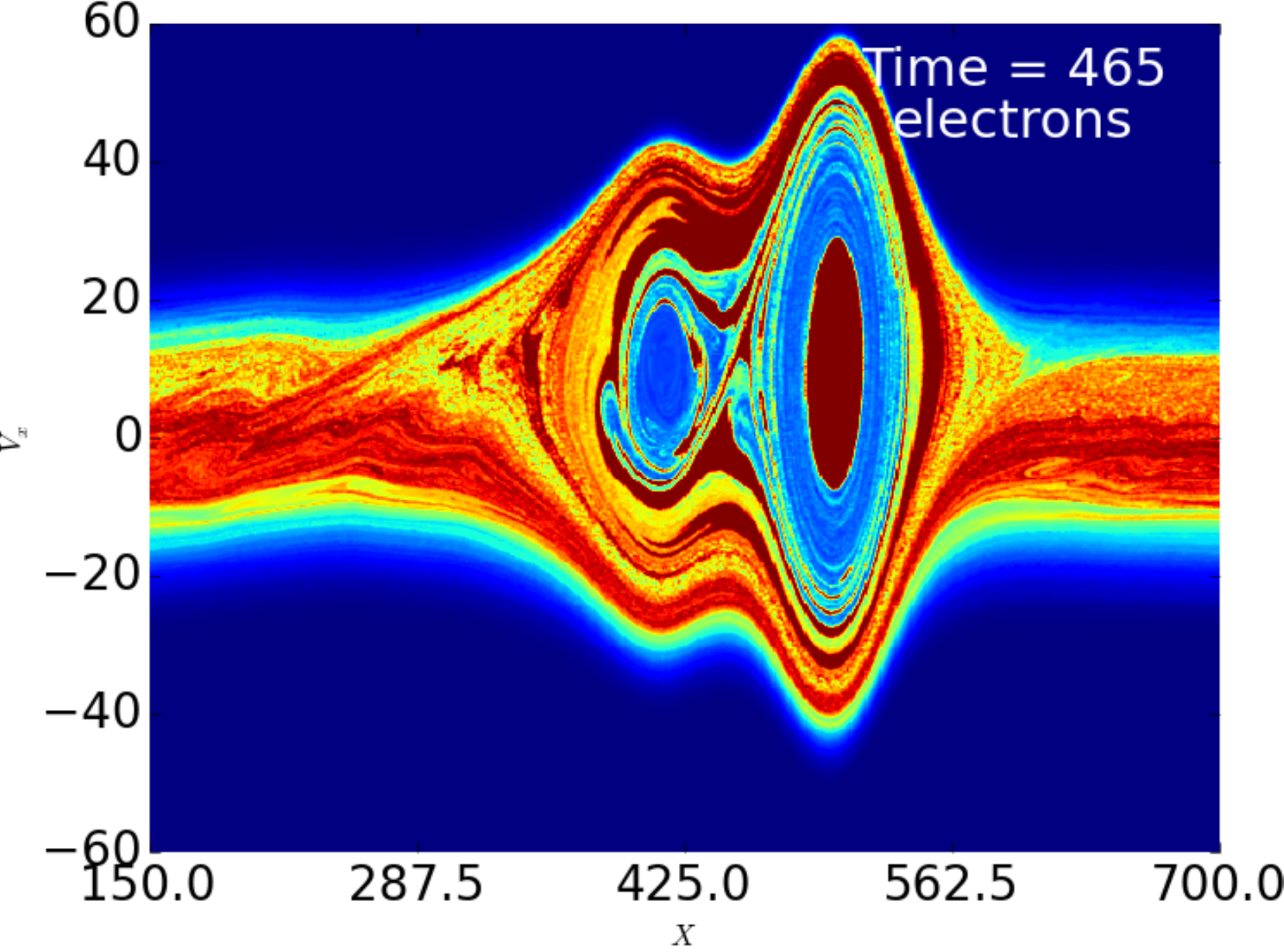}}&
  \subfloat{\includegraphics[width=0.25\textwidth]{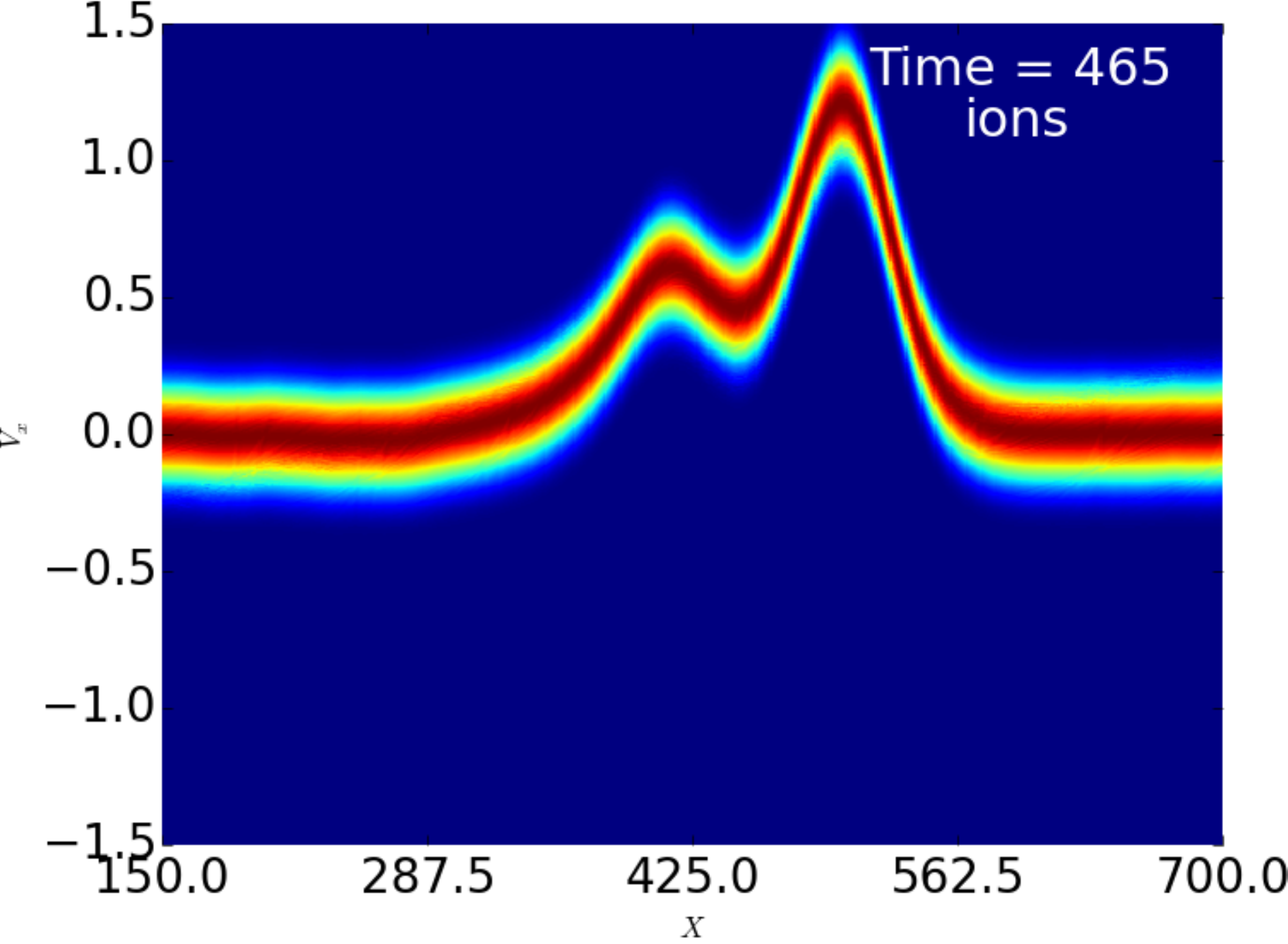}}\\
\end{tabular}
  \caption{The details of the first overtaking collision are shown in the phase space
  for two solitons with different trapping parameters, i.e. $\beta = -0.1$ (hole) and $\beta = 0$ (plateau) 
  in the frame moving with their average velocity. 
  Overlapping of the electrons trapped population on the facing side signals the overlapping of the 
  solitons as seen in Fig. \ref{Fig_NlR_ZsR_num_details}.
  The two IASs exchange some parts of their trapped electron population. 
  They interchange their cores with each other while the outer levels stay the same.}
  \label{Fig_NlR_ZsR_PhaseSpace_details}
\end{figure}

Kinetic details of the overtaking collision
are presented in Fig. \ref{Fig_NlR_ZsR_PhaseSpace_details}. 
The same phenomenon as on the fluid level (Fig. \ref{Fig_NlR_ZsR_num_details}) 
can be seen here as well. 

Firstly, during the collision, the two trapped electron population (here a plateau and a hole)
starts to overlap and lose their boundaries on the facing side. 
As collision continues, more layers of their outer part are exchanged between them. 
Hence, the larger/smaller soliton shrinks/grows on the velocity direction which results
in losing/gaining velocity. 
In the midst of the collision, 
they have reached the same velocity 
as well as the same width in the velocity and spatial directions ($\tau = 425$). 
Therefore they can't continue increasing their overlapping region 
since they are both traveling with the same velocity. 
As the processes continue, the growth/decline in velocity of smaller/larger solitons cause them to 
depart. 
Since during collision they don't exchange their core part of trapped electrons,
hence the smaller/larger soliton splits while having its own core with the outer layer of the larger/smaller soliton. 
We have carried out these types of simulations for solitons with different trapping parameter ($\beta$)
overtaking each other and the same patterns have been witnessed.

Secondly, electron holes/plateaus/humps accompanying the solitons after the overlapping,
split and continue
propagating steadily accompanying their associated solitons. 
However this is in contrast of the behavior of electron holes seen and predicted 
in other phenomena such as 
beam-plasma interactions \cite{berk1970phase,omura1996electron} 
and nonlinear Landau damping
(Bernstein-Greene-Kruskal -BGK- modes)\cite{ghizzo1988stability}.
In these case, it is observed that the electron holes merge in pair until 
the system reaches the stable state of one hole. 
In other words in a periodically bounded system of a plasma,
electron holes tend to merge until one big hole remains. 
Our simulations show that electron holes accompanying IASs
don't show any tendency of merging, even-though they are overlapping during collision. 
Moreover, they split after the collision which has never been reported (to the best of our knowledge) 
in the context of electron holes study.

Furthermore, based on a theory developed by 
Krasovsky \textit{et al.} \cite{krasovsky1999interaction,krasovsky1999interaction_small,krasovsky2003electrostatic} 
utilizing energy conservation principle, 
collision between two electron holes is a dissipative process.
The internal energy of a hole 
(kinetic energy of the trapped electrons in the co-moving frame) 
grows during collision and hence holes warm up. 
This is an irreversible process and causes an effective friction in the energy balance.
In other words any collision between two electron holes should be inelastic and 
they should ultimately merge into one hole. 
Our simulation results display a complete opposite process. 
Krasovsky \textit{et al.} argue that for two holes to merge,
the relative velocity of the holes should be slow enough
so that the trapped electrons oscillate at least once
during collision (condition of merging). 
Here despite the slow relative velocity the two holes \textbf{do not merge}.

\begin{figure}
    \subfloat{\includegraphics[width=0.23\textwidth]{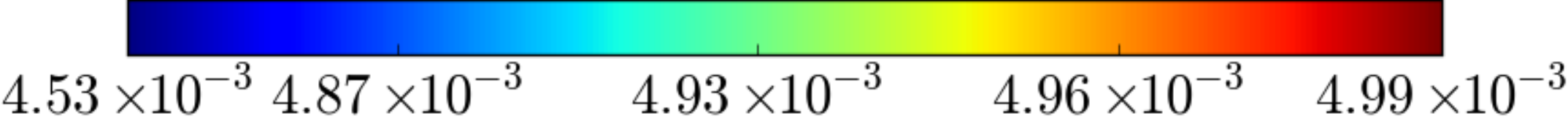}}\ 
    \subfloat{\includegraphics[width=0.23\textwidth]{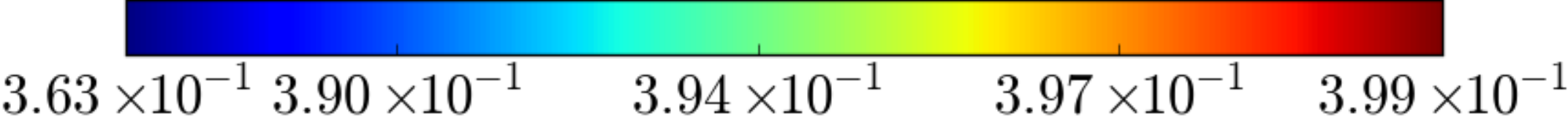}} \\
   
    \subfloat{\includegraphics[width=0.24\textwidth]{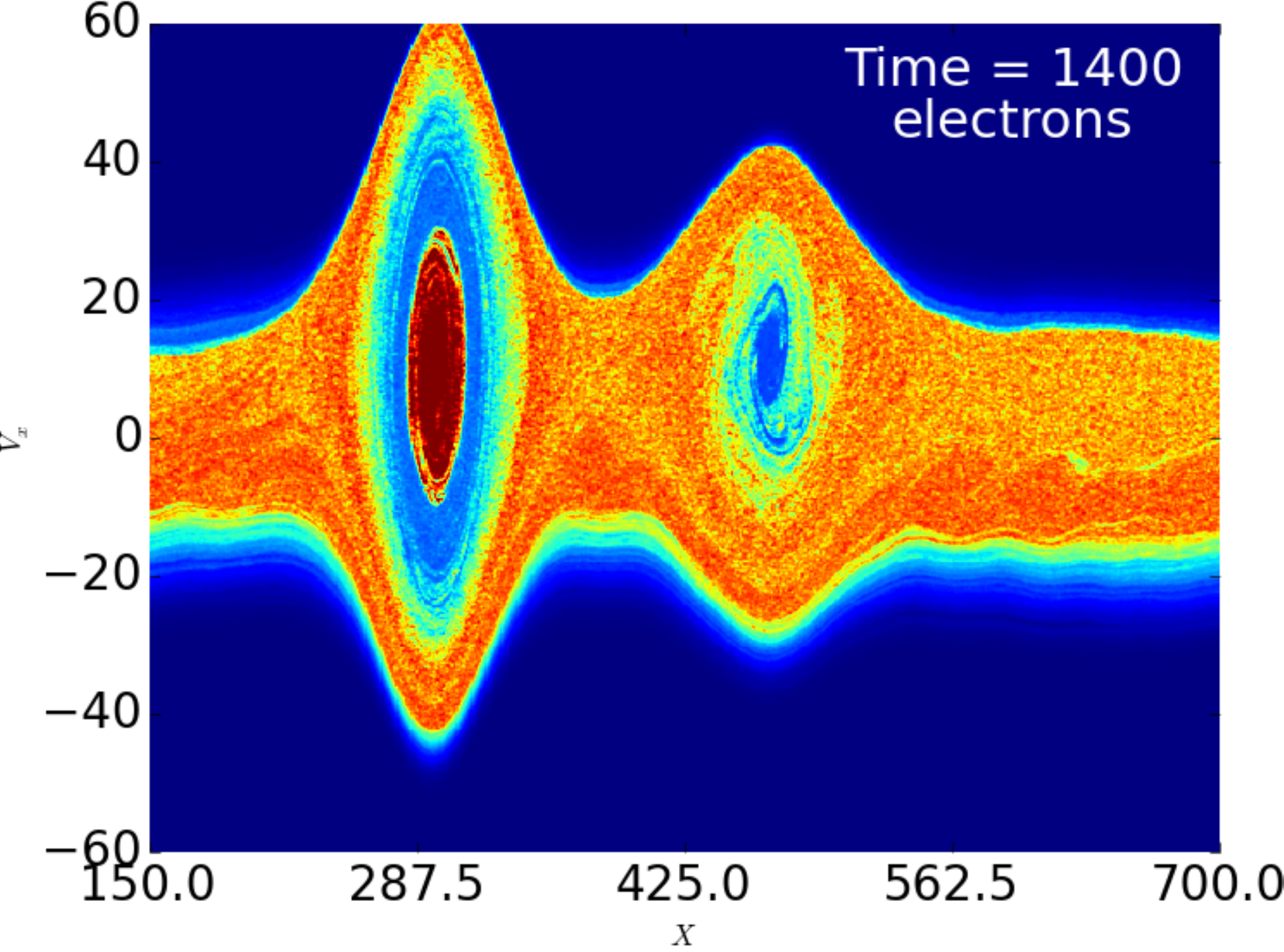}}
    \subfloat{\includegraphics[width=0.24\textwidth]{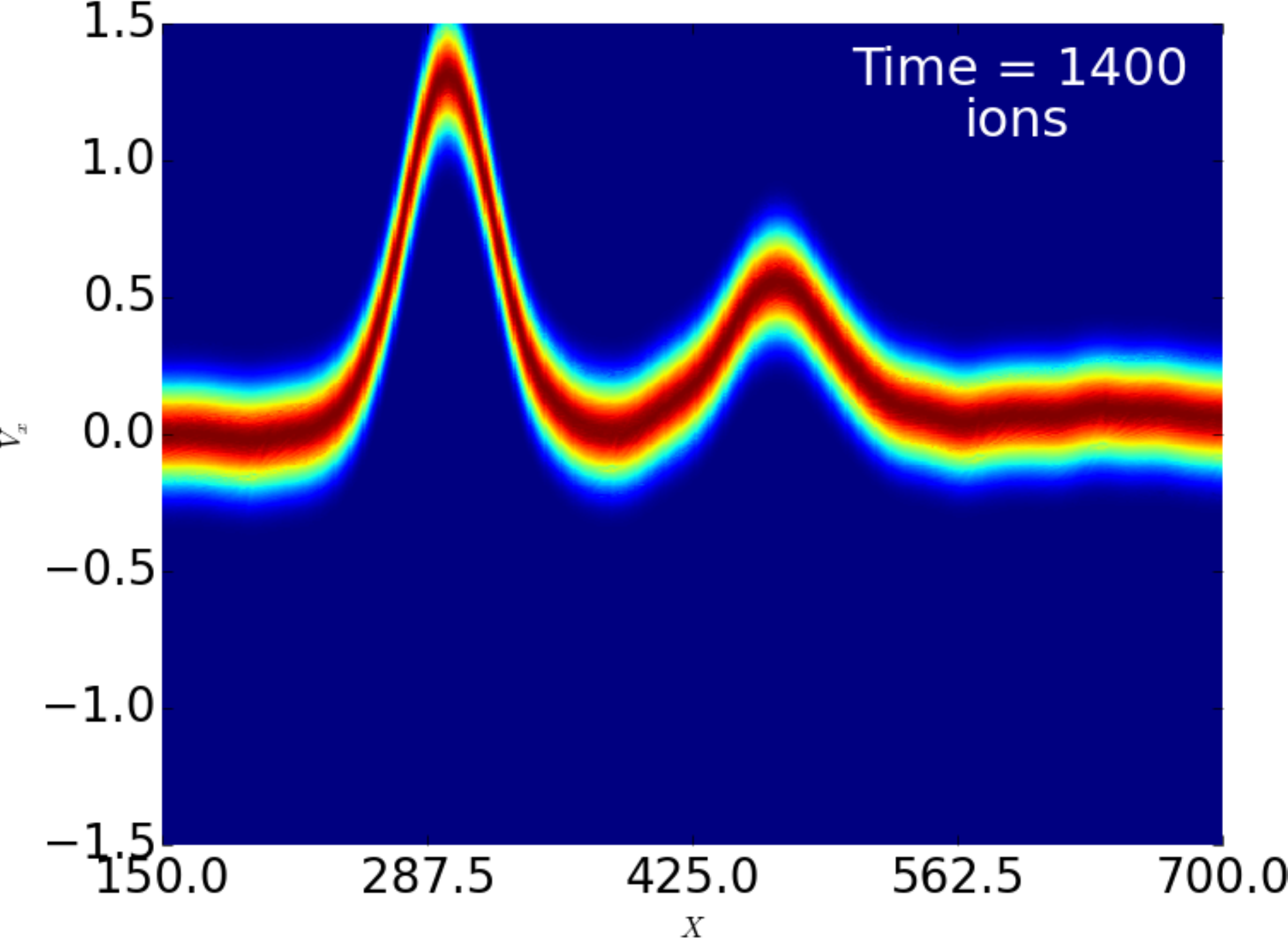}}\\
  
    \subfloat{\includegraphics[width=0.24\textwidth]{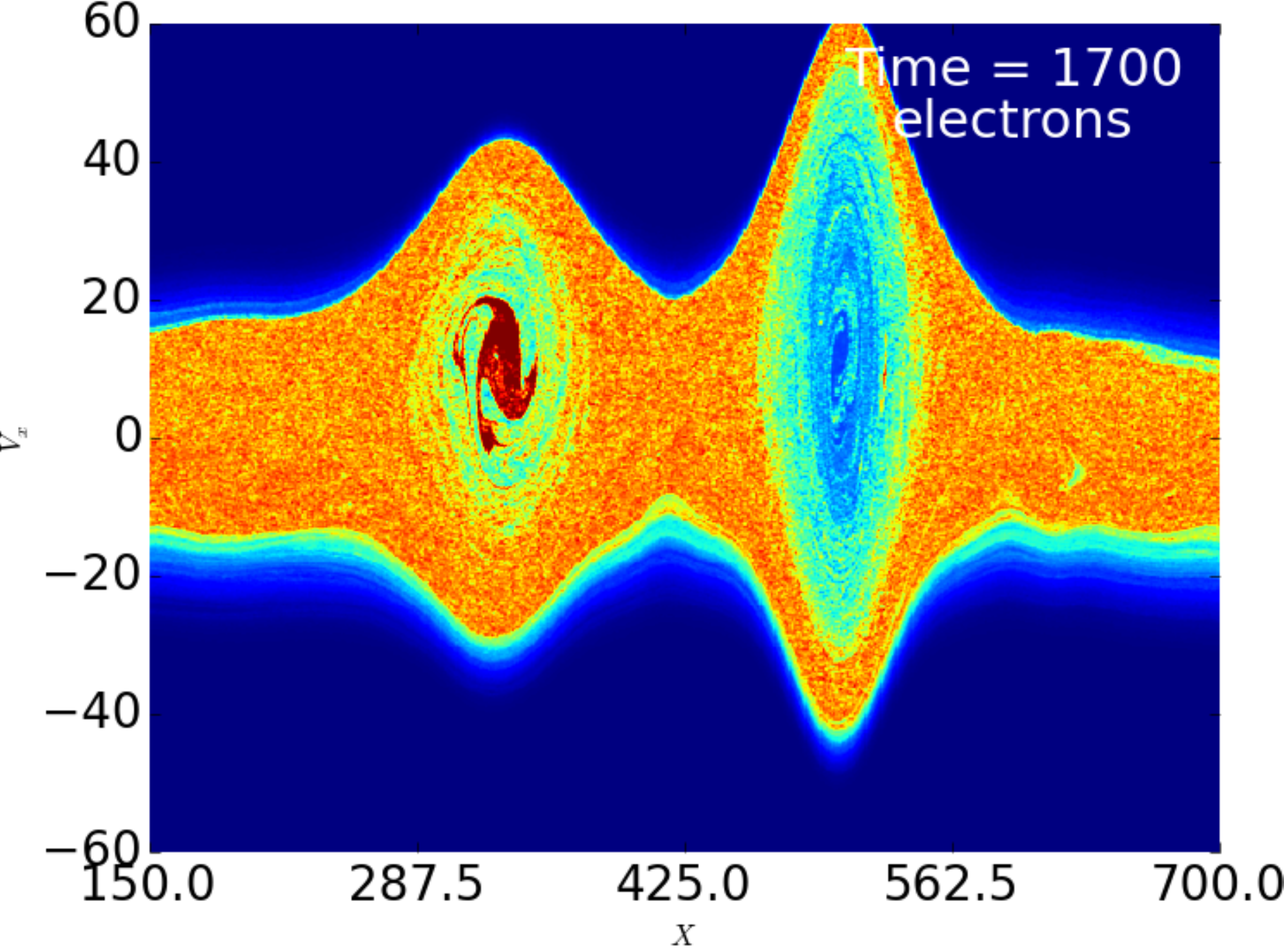}}
    \subfloat{\includegraphics[width=0.24\textwidth]{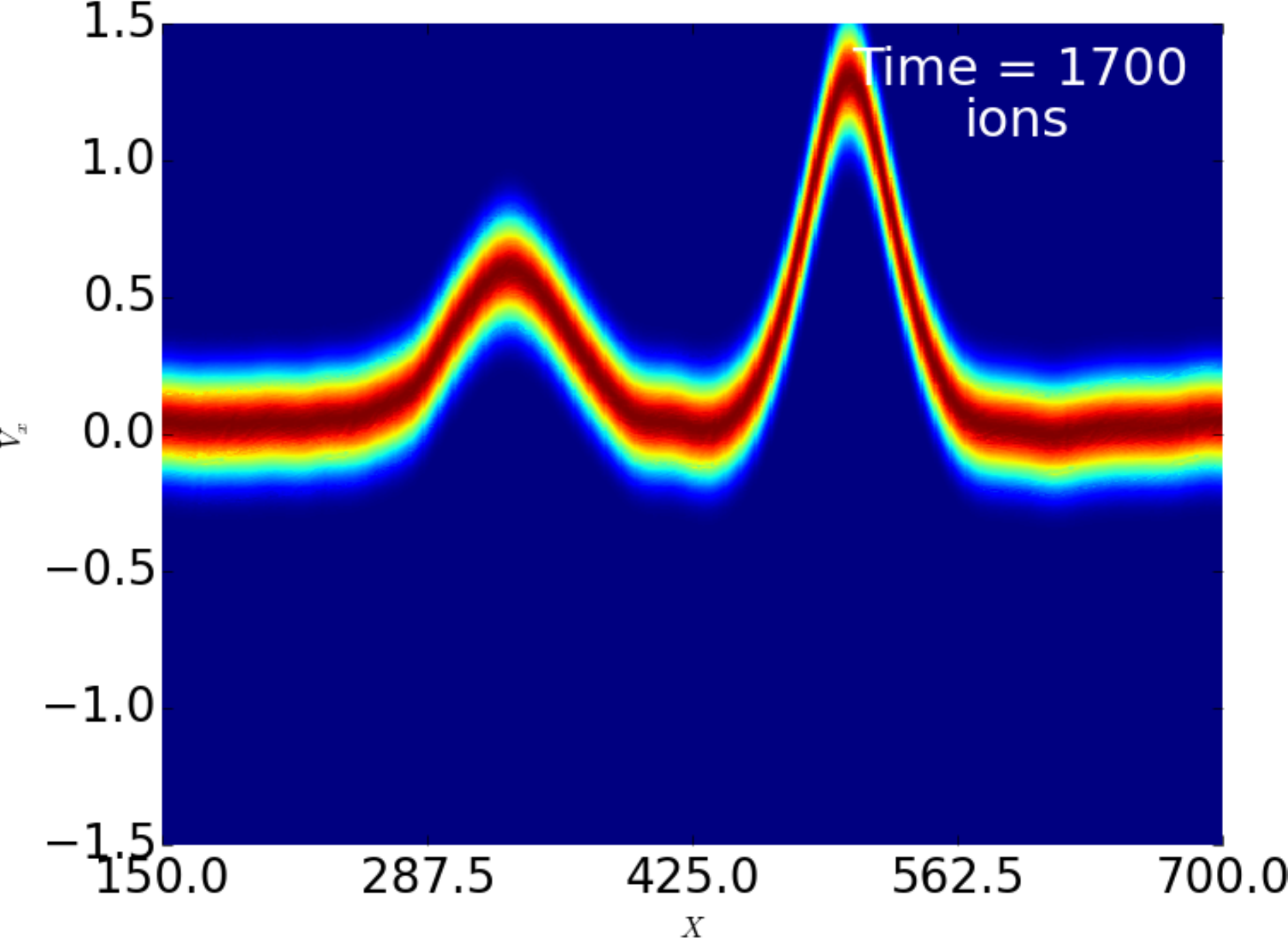}}\\
  \caption{Kinetic details of solitons with $\beta = -0.1$ and $\beta =0$ before and after 
  their second overtaking collision is presented. 
  The same population of trapped electrons as of first overtaking collision (Fig. \ref{Fig_NlR_ZsR_PhaseSpace})
  are exchanged back between them.}
  \label{Fig_NlR_ZsR_PhaseSpace_second}
\end{figure}

By following the temporal evolution even more until the two soliton collide again, 
the same patterns can be witnessed. 
This time, they exchange their outer parts, and 
the resulted trapped population shows the same color as the cores.  
Hence it seems that they exchange the same population 
of the trapped electrons as of the first overtaking collision.
(see Fig. \ref{Fig_NlR_ZsR_PhaseSpace_second}).

\section{Conclusions} \label{Sec_Conclusions}
Three aspects of the ion-acoustic soliton in the presence of trapped electrons
are discussed. 
Initially, it is shown that these nonlinear localized structures can 
propagate for a long time using the simulation method proposed here without 
losing their distinctive characteristics both in fluid and kinetic levels\cite{jenab2016IASWs}. 
This has been used as benchmarking test of the simulation code as well. 

The main focus of the paper is to provide proof of the stability of these IASs 
against mutual overtaking collisions. 
The results on both fluid and kinetic levels are provided and show that the 
physical features such as width, hight, and shape in both spatial and velocity directions
stay the same before and after the collisions. 
Simulations with different trapping parameter, hence different shape of trapped populations, 
prove the stability for a range of the trapping parameter covering negative to positive values. 
However the internal structures of the trapped populations of electrons differs before and after the collisions.

Furthermore, we have studied and presented the details of the collision on both fluid and kinetic level. 
The overall dynamics of overtaking collisions can be reduced into three steps:\\
1) closing-in step\\
2) mid-collision  step\\ 
3) departing step\\
in the first step, the two solitons come close
and start overlapping in the spatial direction. 
Meanwhile, the exchanging of trapped particles starts and  
this cause the fast soliton to lose velocity while the slower one gets faster. 
During this step, due to particle exchange, the amplitude of the larger one
is reduced while the opposite happens
to the smaller soliton. 
Solitons increase their overlapping area until the collision process hits the second step.

At the mid-collision step, which is the exact mid time of collision process, 
the size of the two solitons are equal as well as 
their velocities, hence zero relative velocity.
So they can't continue increasing their overlapping area.
Finally, in the third step, the processes of the first step continue.
The old small-slow soliton becomes new large-fast soliton and visa versa. 
Hence the relative velocity increases and the two solitons starts departing. 
Finally they move further enough that they can't no longer exchange particles 
and therefor the collision process finishes, i.e. overlapping stops. 

Note that the two solitons don't overtake each other, they basically 
come close to each other and exchange particles and depart. 
During this process they exchange their fluid-level identity, i.e. number density profile, 
and this appears as a shift in their trajectory on the fluid level.

In Conclusion, ion-acoustic solitons physical features don not change
on the fluid level before and after overtaking collisions. 
However, on the kinetic level, the internal dynamics of 
the electron trapped population differs. 
But this doe not affect the fluid-level properties, 
hence it is safe to assume ion-acoustic solitons
(in presence of trapping effect of electrons) as solitons, 
structures which can survive collisions.

\acknowledgments
This work is based upon research supported by the National Research Foundation 
and Department of Science and Technology.
Any opinion, findings and conclusions or recommendations expressed in this 
material are those of the authors and therefore the NRF and DST do not accept 
any liability in regard thereto.


\begin{thebibliography}{25}%
\makeatletter
\providecommand \@ifxundefined [1]{%
 \@ifx{#1\undefined}
}%
\providecommand \@ifnum [1]{%
 \ifnum #1\expandafter \@firstoftwo
 \else \expandafter \@secondoftwo
 \fi
}%
\providecommand \@ifx [1]{%
 \ifx #1\expandafter \@firstoftwo
 \else \expandafter \@secondoftwo
 \fi
}%
\providecommand \natexlab [1]{#1}%
\providecommand \enquote  [1]{``#1''}%
\providecommand \bibnamefont  [1]{#1}%
\providecommand \bibfnamefont [1]{#1}%
\providecommand \citenamefont [1]{#1}%
\providecommand \href@noop [0]{\@secondoftwo}%
\providecommand \href [0]{\begingroup \@sanitize@url \@href}%
\providecommand \@href[1]{\@@startlink{#1}\@@href}%
\providecommand \@@href[1]{\endgroup#1\@@endlink}%
\providecommand \@sanitize@url [0]{\catcode `\\12\catcode `\$12\catcode
  `\&12\catcode `\#12\catcode `\^12\catcode `\_12\catcode `\%12\relax}%
\providecommand \@@startlink[1]{}%
\providecommand \@@endlink[0]{}%
\providecommand \url  [0]{\begingroup\@sanitize@url \@url }%
\providecommand \@url [1]{\endgroup\@href {#1}{\urlprefix }}%
\providecommand \urlprefix  [0]{URL }%
\providecommand \Eprint [0]{\href }%
\providecommand \doibase [0]{http://dx.doi.org/}%
\providecommand \selectlanguage [0]{\@gobble}%
\providecommand \bibinfo  [0]{\@secondoftwo}%
\providecommand \bibfield  [0]{\@secondoftwo}%
\providecommand \translation [1]{[#1]}%
\providecommand \BibitemOpen [0]{}%
\providecommand \bibitemStop [0]{}%
\providecommand \bibitemNoStop [0]{.\EOS\space}%
\providecommand \EOS [0]{\spacefactor3000\relax}%
\providecommand \BibitemShut  [1]{\csname bibitem#1\endcsname}%
\let\auto@bib@innerbib\@empty
\bibitem [{\citenamefont {Abbasi}, \citenamefont {Jenab},\ and\ \citenamefont
  {Pajouh}(2011)}]{jenab2011preventing}%
  \BibitemOpen
  \bibfield  {author} {\bibinfo {author} {\bibnamefont {Abbasi}, \bibfnamefont
  {H.}}, \bibinfo {author} {\bibnamefont {Jenab}, \bibfnamefont {M.}}, \ and\
  \bibinfo {author} {\bibnamefont {Pajouh}, \bibfnamefont {H.~H.}},\
  }\href@noop {} {\bibfield  {journal} {\bibinfo  {journal} {Physical Review
  E}\ }\textbf {\bibinfo {volume} {84}},\ \bibinfo {pages} {036702} (\bibinfo
  {year} {2011})}\BibitemShut {NoStop}%
\bibitem [{\citenamefont {Berk}, \citenamefont {Nielsen},\ and\ \citenamefont
  {Roberts}(1970)}]{berk1970phase}%
  \BibitemOpen
  \bibfield  {author} {\bibinfo {author} {\bibnamefont {Berk}, \bibfnamefont
  {H.}}, \bibinfo {author} {\bibnamefont {Nielsen}, \bibfnamefont {C.}}, \ and\
  \bibinfo {author} {\bibnamefont {Roberts}, \bibfnamefont {K.}},\ }\href@noop
  {} {\bibfield  {journal} {\bibinfo  {journal} {Physics of Fluids
  (1958-1988)}\ }\textbf {\bibinfo {volume} {13}},\ \bibinfo {pages} {980}
  (\bibinfo {year} {1970})}\BibitemShut {NoStop}%
\bibitem [{\citenamefont {Bernstein}, \citenamefont {Greene},\ and\
  \citenamefont {Kruskal}(1957)}]{bernstein1957exact}%
  \BibitemOpen
  \bibfield  {author} {\bibinfo {author} {\bibnamefont {Bernstein},
  \bibfnamefont {I.~B.}}, \bibinfo {author} {\bibnamefont {Greene},
  \bibfnamefont {J.~M.}}, \ and\ \bibinfo {author} {\bibnamefont {Kruskal},
  \bibfnamefont {M.~D.}},\ }\href@noop {} {\bibfield  {journal} {\bibinfo
  {journal} {Physical Review}\ }\textbf {\bibinfo {volume} {108}},\ \bibinfo
  {pages} {546} (\bibinfo {year} {1957})}\BibitemShut {NoStop}%
\bibitem [{\citenamefont {Ghizzo}\ \emph {et~al.}(1988)\citenamefont {Ghizzo},
  \citenamefont {Izrar}, \citenamefont {Bertrand}, \citenamefont {Fijalkow},
  \citenamefont {Feix},\ and\ \citenamefont {Shoucri}}]{ghizzo1988stability}%
  \BibitemOpen
  \bibfield  {author} {\bibinfo {author} {\bibnamefont {Ghizzo}, \bibfnamefont
  {A.}}, \bibinfo {author} {\bibnamefont {Izrar}, \bibfnamefont {B.}}, \bibinfo
  {author} {\bibnamefont {Bertrand}, \bibfnamefont {P.}}, \bibinfo {author}
  {\bibnamefont {Fijalkow}, \bibfnamefont {E.}}, \bibinfo {author}
  {\bibnamefont {Feix}, \bibfnamefont {M.}}, \ and\ \bibinfo {author}
  {\bibnamefont {Shoucri}, \bibfnamefont {M.}},\ }\href@noop {} {\bibfield
  {journal} {\bibinfo  {journal} {Physics of Fluids (1958-1988)}\ }\textbf
  {\bibinfo {volume} {31}},\ \bibinfo {pages} {72} (\bibinfo {year}
  {1988})}\BibitemShut {NoStop}%
\bibitem [{\citenamefont {Ikezi}, \citenamefont {Taylor},\ and\ \citenamefont
  {Baker}(1970)}]{ikezi1970formation}%
  \BibitemOpen
  \bibfield  {author} {\bibinfo {author} {\bibnamefont {Ikezi}, \bibfnamefont
  {H.}}, \bibinfo {author} {\bibnamefont {Taylor}, \bibfnamefont {R.}}, \ and\
  \bibinfo {author} {\bibnamefont {Baker}, \bibfnamefont {D.}},\ }\href@noop {}
  {\bibfield  {journal} {\bibinfo  {journal} {Physical Review Letters}\
  }\textbf {\bibinfo {volume} {25}},\ \bibinfo {pages} {11} (\bibinfo {year}
  {1970})}\BibitemShut {NoStop}%
\bibitem [{\citenamefont {Jenab}\ and\ \citenamefont
  {Spanier}(2016)}]{jenab2016IASWs}%
  \BibitemOpen
  \bibfield  {author} {\bibinfo {author} {\bibnamefont {Jenab}, \bibfnamefont
  {S.~H.}}\ and\ \bibinfo {author} {\bibnamefont {Spanier}, \bibfnamefont
  {F.}},\ }\href@noop {} {\bibfield  {journal} {\bibinfo  {journal} {Physics of
  Plasmas}\ }\textbf {\bibinfo {volume} {23}},\ \bibinfo {pages} {102306}
  (\bibinfo {year} {2016})}\BibitemShut {NoStop}%
\bibitem [{\citenamefont {Kakad}, \citenamefont {Omura},\ and\ \citenamefont
  {Kakad}(2013)}]{Kakad2013}%
  \BibitemOpen
  \bibfield  {author} {\bibinfo {author} {\bibnamefont {Kakad}, \bibfnamefont
  {A.}}, \bibinfo {author} {\bibnamefont {Omura}, \bibfnamefont {Y.}}, \ and\
  \bibinfo {author} {\bibnamefont {Kakad}, \bibfnamefont {B.}},\ }\href@noop {}
  {\bibfield  {journal} {\bibinfo  {journal} {Physics of Plasmas
  (1994-present)}\ }\textbf {\bibinfo {volume} {20}},\ \bibinfo {pages}
  {062103} (\bibinfo {year} {2013})}\BibitemShut {NoStop}%
\bibitem [{\citenamefont {Kakad}, \citenamefont {Kakad},\ and\ \citenamefont
  {Omura}(2014)}]{Kakad20145589}%
  \BibitemOpen
  \bibfield  {author} {\bibinfo {author} {\bibnamefont {Kakad}, \bibfnamefont
  {B.}}, \bibinfo {author} {\bibnamefont {Kakad}, \bibfnamefont {A.}}, \ and\
  \bibinfo {author} {\bibnamefont {Omura}, \bibfnamefont {Y.}},\ }\href@noop {}
  {\bibfield  {journal} {\bibinfo  {journal} {Journal of Geophysical Research:
  Space Physics}\ }\textbf {\bibinfo {volume} {119}},\ \bibinfo {pages} {5589}
  (\bibinfo {year} {2014})}\BibitemShut {NoStop}%
\bibitem [{\citenamefont {Kazeminezhad}, \citenamefont {Kuhn},\ and\
  \citenamefont {Tavakoli}(2003)}]{kazeminezhad2003vlasov}%
  \BibitemOpen
  \bibfield  {author} {\bibinfo {author} {\bibnamefont {Kazeminezhad},
  \bibfnamefont {F.}}, \bibinfo {author} {\bibnamefont {Kuhn}, \bibfnamefont
  {S.}}, \ and\ \bibinfo {author} {\bibnamefont {Tavakoli}, \bibfnamefont
  {A.}},\ }\href@noop {} {\bibfield  {journal} {\bibinfo  {journal} {Physical
  Review E}\ }\textbf {\bibinfo {volume} {67}},\ \bibinfo {pages} {026704}
  (\bibinfo {year} {2003})}\BibitemShut {NoStop}%
\bibitem [{\citenamefont {Krasovsky}, \citenamefont {Matsumoto},\ and\
  \citenamefont {Omura}(1999{\natexlab{a}})}]{krasovsky1999interaction}%
  \BibitemOpen
  \bibfield  {author} {\bibinfo {author} {\bibnamefont {Krasovsky},
  \bibfnamefont {V.}}, \bibinfo {author} {\bibnamefont {Matsumoto},
  \bibfnamefont {H.}}, \ and\ \bibinfo {author} {\bibnamefont {Omura},
  \bibfnamefont {Y.}},\ }\href@noop {} {\bibfield  {journal} {\bibinfo
  {journal} {Nonlinear Processes in Geophysics}\ }\textbf {\bibinfo {volume}
  {6}},\ \bibinfo {pages} {205} (\bibinfo {year}
  {1999}{\natexlab{a}})}\BibitemShut {NoStop}%
\bibitem [{\citenamefont {Krasovsky}, \citenamefont {Matsumoto},\ and\
  \citenamefont {Omura}(1999{\natexlab{b}})}]{krasovsky1999interaction_small}%
  \BibitemOpen
  \bibfield  {author} {\bibinfo {author} {\bibnamefont {Krasovsky},
  \bibfnamefont {V.}}, \bibinfo {author} {\bibnamefont {Matsumoto},
  \bibfnamefont {H.}}, \ and\ \bibinfo {author} {\bibnamefont {Omura},
  \bibfnamefont {Y.}},\ }\href@noop {} {\bibfield  {journal} {\bibinfo
  {journal} {Physica Scripta}\ }\textbf {\bibinfo {volume} {60}},\ \bibinfo
  {pages} {438} (\bibinfo {year} {1999}{\natexlab{b}})}\BibitemShut {NoStop}%
\bibitem [{\citenamefont {Krasovsky}, \citenamefont {Matsumoto},\ and\
  \citenamefont {Omura}(2003)}]{krasovsky2003electrostatic}%
  \BibitemOpen
  \bibfield  {author} {\bibinfo {author} {\bibnamefont {Krasovsky},
  \bibfnamefont {V.}}, \bibinfo {author} {\bibnamefont {Matsumoto},
  \bibfnamefont {H.}}, \ and\ \bibinfo {author} {\bibnamefont {Omura},
  \bibfnamefont {Y.}},\ }\href@noop {} {\bibfield  {journal} {\bibinfo
  {journal} {Journal of Geophysical Research: Space Physics}\ }\textbf
  {\bibinfo {volume} {108}} (\bibinfo {year} {2003})}\BibitemShut {NoStop}%
\bibitem [{\citenamefont {Kuznetsov}, \citenamefont {Rubenchik},\ and\
  \citenamefont {Zakharov}(1986)}]{kuznetsov1986soliton}%
  \BibitemOpen
  \bibfield  {author} {\bibinfo {author} {\bibnamefont {Kuznetsov},
  \bibfnamefont {E.}}, \bibinfo {author} {\bibnamefont {Rubenchik},
  \bibfnamefont {A.}}, \ and\ \bibinfo {author} {\bibnamefont {Zakharov},
  \bibfnamefont {V.~E.}},\ }\href@noop {} {\bibfield  {journal} {\bibinfo
  {journal} {Physics Reports}\ }\textbf {\bibinfo {volume} {142}},\ \bibinfo
  {pages} {103} (\bibinfo {year} {1986})}\BibitemShut {NoStop}%
\bibitem [{\citenamefont {Lonngren}(1983)}]{lonngren1983soliton}%
  \BibitemOpen
  \bibfield  {author} {\bibinfo {author} {\bibnamefont {Lonngren},
  \bibfnamefont {K.~E.}},\ }\href@noop {} {\bibfield  {journal} {\bibinfo
  {journal} {Plasma Physics}\ }\textbf {\bibinfo {volume} {25}},\ \bibinfo
  {pages} {943} (\bibinfo {year} {1983})}\BibitemShut {NoStop}%
\bibitem [{\citenamefont {Nunn}(1993)}]{nunn1993novel}%
  \BibitemOpen
  \bibfield  {author} {\bibinfo {author} {\bibnamefont {Nunn}, \bibfnamefont
  {D.}},\ }\href@noop {} {\bibfield  {journal} {\bibinfo  {journal} {Journal of
  Computational Physics}\ }\textbf {\bibinfo {volume} {108}},\ \bibinfo {pages}
  {180} (\bibinfo {year} {1993})}\BibitemShut {NoStop}%
\bibitem [{\citenamefont {Omura}\ \emph {et~al.}(1996)\citenamefont {Omura},
  \citenamefont {Matsumoto}, \citenamefont {Miyake},\ and\ \citenamefont
  {Kojima}}]{omura1996electron}%
  \BibitemOpen
  \bibfield  {author} {\bibinfo {author} {\bibnamefont {Omura}, \bibfnamefont
  {Y.}}, \bibinfo {author} {\bibnamefont {Matsumoto}, \bibfnamefont {H.}},
  \bibinfo {author} {\bibnamefont {Miyake}, \bibfnamefont {T.}}, \ and\
  \bibinfo {author} {\bibnamefont {Kojima}, \bibfnamefont {H.}},\ }\href@noop
  {} {\bibfield  {journal} {\bibinfo  {journal} {Journal of Geophysical
  Research: Space Physics}\ }\textbf {\bibinfo {volume} {101}},\ \bibinfo
  {pages} {2685} (\bibinfo {year} {1996})}\BibitemShut {NoStop}%
\bibitem [{\citenamefont {Qi}\ \emph {et~al.}(2015)\citenamefont {Qi},
  \citenamefont {Xu}, \citenamefont {Zhao}, \citenamefont {Zhang},
  \citenamefont {Duan},\ and\ \citenamefont {Yang}}]{Qi20153815}%
  \BibitemOpen
  \bibfield  {author} {\bibinfo {author} {\bibnamefont {Qi}, \bibfnamefont
  {X.}}, \bibinfo {author} {\bibnamefont {Xu}, \bibfnamefont {Y.-X.}}, \bibinfo
  {author} {\bibnamefont {Zhao}, \bibfnamefont {X.-Y.}}, \bibinfo {author}
  {\bibnamefont {Zhang}, \bibfnamefont {L.-Y.}}, \bibinfo {author}
  {\bibnamefont {Duan}, \bibfnamefont {W.-S.}}, \ and\ \bibinfo {author}
  {\bibnamefont {Yang}, \bibfnamefont {L.}},\ }\href@noop {} {\bibfield
  {journal} {\bibinfo  {journal} {IEEE Transactions on Plasma Science}\
  }\textbf {\bibinfo {volume} {43}},\ \bibinfo {pages} {3815} (\bibinfo {year}
  {2015})}\BibitemShut {NoStop}%
\bibitem [{\citenamefont {Sagdeev}(1966)}]{Sagdeev}%
  \BibitemOpen
  \bibfield  {author} {\bibinfo {author} {\bibnamefont {Sagdeev}, \bibfnamefont
  {R.}},\ }\href@noop {} {\bibfield  {journal} {\bibinfo  {journal} {Reviews of
  Plasma Physics}\ }\textbf {\bibinfo {volume} {4}},\ \bibinfo {pages} {23}
  (\bibinfo {year} {1966})}\BibitemShut {NoStop}%
\bibitem [{\citenamefont {Schamel}(1971)}]{schamel_1}%
  \BibitemOpen
  \bibfield  {author} {\bibinfo {author} {\bibnamefont {Schamel}, \bibfnamefont
  {H.}},\ }\href@noop {} {\bibfield  {journal} {\bibinfo  {journal} {Plasma
  Physics}\ }\textbf {\bibinfo {volume} {13}},\ \bibinfo {pages} {491}
  (\bibinfo {year} {1971})}\BibitemShut {NoStop}%
\bibitem [{\citenamefont {Schamel}(1972)}]{schamel_3}%
  \BibitemOpen
  \bibfield  {author} {\bibinfo {author} {\bibnamefont {Schamel}, \bibfnamefont
  {H.}},\ }\href@noop {} {\bibfield  {journal} {\bibinfo  {journal} {Plasma
  Physics}\ }\textbf {\bibinfo {volume} {14}},\ \bibinfo {pages} {905}
  (\bibinfo {year} {1972})}\BibitemShut {NoStop}%
\bibitem [{\citenamefont {Shafranov}(2012)}]{shafranov2012reviews}%
  \BibitemOpen
  \bibfield  {author} {\bibinfo {author} {\bibnamefont {Shafranov},
  \bibfnamefont {V.~D.}},\ }\href@noop {} {\emph {\bibinfo {title} {Reviews of
  Plasma Physics}}},\ Vol.~\bibinfo {volume} {22}\ (\bibinfo  {publisher}
  {Springer Science \& Business Media},\ \bibinfo {year} {2012})\BibitemShut
  {NoStop}%
\bibitem [{\citenamefont {Sharma}, \citenamefont {Sengupta},\ and\
  \citenamefont {Sen}(2015)}]{Sharma2015}%
  \BibitemOpen
  \bibfield  {author} {\bibinfo {author} {\bibnamefont {Sharma}, \bibfnamefont
  {S.}}, \bibinfo {author} {\bibnamefont {Sengupta}, \bibfnamefont {S.}}, \
  and\ \bibinfo {author} {\bibnamefont {Sen}, \bibfnamefont {A.}},\ }\href@noop
  {} {\bibfield  {journal} {\bibinfo  {journal} {Physics of Plasmas
  (1994-present)}\ }\textbf {\bibinfo {volume} {22}},\ \bibinfo {pages}
  {022115} (\bibinfo {year} {2015})}\BibitemShut {NoStop}%
\bibitem [{\citenamefont {Tran}(1979)}]{tran1979ion}%
  \BibitemOpen
  \bibfield  {author} {\bibinfo {author} {\bibnamefont {Tran}, \bibfnamefont
  {M.}},\ }\href@noop {} {\bibfield  {journal} {\bibinfo  {journal} {Physica
  Scripta}\ }\textbf {\bibinfo {volume} {20}},\ \bibinfo {pages} {317}
  (\bibinfo {year} {1979})}\BibitemShut {NoStop}%
\bibitem [{\citenamefont {Wadati}(2001)}]{Wadati2001841}%
  \BibitemOpen
  \bibfield  {author} {\bibinfo {author} {\bibnamefont {Wadati}, \bibfnamefont
  {M.}},\ }\href@noop {} {\bibfield  {journal} {\bibinfo  {journal} {Pramana}\
  }\textbf {\bibinfo {volume} {57}},\ \bibinfo {pages} {841} (\bibinfo {year}
  {2001})}\BibitemShut {NoStop}%
\bibitem [{\citenamefont {Zabusky}\ and\ \citenamefont
  {Kruskal}(1965)}]{Zabusky1965}%
  \BibitemOpen
  \bibfield  {author} {\bibinfo {author} {\bibnamefont {Zabusky}, \bibfnamefont
  {N.~J.}}\ and\ \bibinfo {author} {\bibnamefont {Kruskal}, \bibfnamefont
  {M.~D.}},\ }\href@noop {} {\bibfield  {journal} {\bibinfo  {journal}
  {Physical review letters}\ }\textbf {\bibinfo {volume} {15}},\ \bibinfo
  {pages} {240} (\bibinfo {year} {1965})}\BibitemShut {NoStop}%
\end{thebibliography}

%

\end{document}